\documentclass[preprint,preprintnumbers,amsmath,amssymb,nofootinbib]{revtex4}
\usepackage{graphicx}
\usepackage{float}
\usepackage{color}
\newcommand\nn{\nonumber \\}
\newcommand\vk{{\vec{k}}}

\newcommand\dert{{\frac{\rm d}{{\rm d}t} }}

\newcommand{\lk}{\left(}
\newcommand{\rk}{\right)}
\newcommand{\ltk}{\left\{}
\newcommand{\rtk}{\right\}}
\newcommand{\ldk}{\left[}

\newcommand{\rdk}{\right]}
\newcommand\beq{ \begin{eqnarray} }
\newcommand\eeq{ \end{eqnarray} }

\begin{document}



\title{Transport coefficients of $O(N)$ scalar field theories 
close to the critical point} 

\author{Eiji Nakano}
\affiliation{%
Physics Division, Faculty of Science, Kochi University, Kochi 780-8520, Japan, and \\
Extreme Matter institute at GSI Helmholtzzentrum f\"ur Schwerionenforschung, 
D-64291 Darmstadt, Germany}

\author{Vladimir Skokov}
\affiliation{%
  GSI Helmholtzzentrum f\"ur Schwerionenforschung, D-64291 Darmstadt,
  Germany}

\author{Bengt Friman}
\affiliation{%
  GSI Helmholtzzentrum f\"ur Schwerionenforschung, D-64291 Darmstadt,
  Germany}


\begin{abstract}
We investigate the critical dynamics  of $O(N)$-symmetric 
scalar field theories to determine the critical exponents of transport coefficients 
as a second-order phase transition is approached from the symmetric phase. 
A set of stochastic equations of motion for the slow modes is formulated, 
and the long wavelength dynamics is examined for an arbitrary number of field components, 
$N$, in the framework of the dynamical renormalization group within the $\varepsilon$ expansion. 
We find that for a single component scalar field theory, $N=1$, 
the system reduces to the model C of critical dynamics, 
whereas for $N>1$ the model G is effectively restored 
owing to dominance of $O(N)$-symmetric charge fluctuations. 
In both cases, the shear viscosity remains finite in the critical region. 
On the other hand, we find that 
the bulk viscosity diverges 
as the correlation length squared, for $N=1$, 
while it remains finite for $N>1$. 

\end{abstract}


\maketitle
\section{Introduction}
In recent decades transport coefficients of the quantum chromodynamics (QCD) 
have attracted much interest 
in the context of Relativistic Heavy Ion Collider (RHIC) experiment, 
which aims at creating and studying a quark-gluon plasma. 
One of the interesting findings emerging from the experimental program at RHIC, 
the large elliptic flow $v_2$ observed in high energy non-central collisions, implies that 
the spatial anisotropy of the initial state created in the collision is efficiently converted during the expansion 
 to a transverse momentum anisotropy of the observed 
hadrons~\cite{Ackermann:2000tr,Adler:2003kt,Adams:2003am}. 
These experimental results 
are well described by ideal hydrodynamics 
with vanishing viscosity \cite{Shuryak:2003xe,Teaney:2003kp,Romatschke:2007mq}. 
Thus, the large elliptic flow observed in such collisions implies that the matter created in collisions 
behaves as an almost perfect fluid. 

Although the transport coefficients in viscous hydrodynamics are phenomenological parameters, 
they can, in principle, be computed from a microscopic theory. 
Since the shear viscosity, one of the transport coefficients, 
has a direct influence on the elliptic flow, 
the experimental results have triggered numerous theoretical efforts to unravel its behavior 
as a function of thermodynamic variables. In general, these are performed in the framework of kinetic theory, e.g. 
using the Boltzmann equation, applied to effective theories of QCD 
\cite{Gavin:1985ph,Prakash:1993bt,Davesne:1995ms,Dobado:2003wr,Chen:2006iga,Chen:2007xe,Chen:2007kx,Sasaki:2008fg}
and to perturbative QCD \cite{Arnold:2000dr,Arnold:2003zc,Xu:2007jv,Chen:2009sm,Niemi:2011ix}.  
Furthermore, some results on the temperature dependence of the transport coefficients have been obtained in 
lattice simulations~\cite{Karsch:1986cq,Nakamura:2004sy,Meyer:2007ic,Meyer:2007dy,Karsch:2007jc,Kharzeev:2007wb,Huebner:2008as}. 

The results of the RHIC experiments have motivated recent work on a field theoretical approach 
to evaluate transport coefficients. 
The $O(N)$ scalar field theory offers a testing ground for developing computational methods
before facing the complications of a full QCD calculation. The scalar field theory has in fact long been studied 
as a prototype theory in many contexts of physics. 
Moreover, for $N=4$, the $O(N)$ model serve as  
 a low energy chiral effective theory for two flavor QCD 
\cite{Pisarski:1983ms}. 
A general Lagrangian density for the $O(N)$ scalar field theory is given by 
\beq
{\mathcal L}=\frac{1}{2} \lk \partial \phi_i\rk ^2 -\frac{1}{2} r \phi_i^2
-\frac{1}{4} u \lk \phi_i^2 \rk^2, 
\eeq
where $r$ is the mass parameter, $u$ is the coupling constant, 
and the implicit summation over $i$ runs from $i=1$ to $N$. 
Recently, the dynamical properties of the $O(N)$ scalar theory, 
in particular behavior of the transport coefficients, 
formulated microscopically in the Green-Kubo-Nakano 
linear response theory, have been explored in several theoretical studies. 

The shear viscosity, $\eta$, of the scalar field theory was first studied 
by Hosoya {\it et. al.}~\cite{Hosoya:1983id} and
Jeon and Yaffe~\cite{Jeon:1994if,Jeon:1995zm}  
in thermal field theory.
Later on, the large $N$ behavior was examined by Aarts and Resco~\cite{Aarts:2003bk,Aarts:2004sd}. 
These calculations demonstrated that $\eta$ is an increasing function of temperature, 
$T$. At high temperatures, 
\beq
\eta \sim \frac{N^2}{u^2}T^3. 
\label{highT}
\eeq
The cubic power in temperature can be understood on dimensional grounds, 
and the factor $N^2$ is attributed to the scaling of the coupling constant $u$ with $1/N$. 
The inverse power of the coupling constant in Eq.~(\ref{highT}) implies
that the shear viscosity is a non-perturbative quantity. The precise numerical factor 
in Eq.~(\ref{highT}) can be obtained 
by a resummation of ladder type diagrams. 
In Refs.~\cite{Jeon:1994if,Jeon:1995zm}, 
it was found that 
the ladder resummation is equivalent to 
the linearized Boltzmann equation with a thermal mass term. 
Some systematic approaches for computing higher order corrections are presented in Refs. \cite{York:2008rr,Carrington:2009kh,Hidaka:2010gh}, 
and relevant issues on the transport equation are discussed in Refs. \cite{Carrington:2004tm,Berges:2005md}.

In the present work, we discuss the critical behavior of the shear viscosity 
and other transport coefficients in the $O(N)$ scalar field theory.  
As demonstrated by Wilson using the renormalization group approach, 
there is a second-order phase transition 
in the $O(N)$ scalar field theory. 
Kinetic approaches employed for computing the transport coefficients 
(see e.g.~\cite{Chen:2007jq,Chen:2010vg}
and a discussion in Ref.~\cite{Schafer:2009dj}) 
rely heavily on Boltzmann-like 
approximations, which take only the single particle distribution into account 
and neglect higher order correlations. Although these correlations, may be unimportant far
from the critical point, they play an important role in the critical region.  

In our study of the critical transport properties, 
we employ the dynamical renormalization group (DRG) combined with the epsilon expansion 
\footnote{An alternative non-perturbative approach to the epsilon expansion would 
be a direct application of the functional renormalization group 
(see e.g. Ref.~\cite{Berges:2000ew} for a review)
to a  quantum-field model  constructed to be equivalent to the stochastic equations of motion~\cite{Janssen,DeDominicis:1977fw}. 
This  method was tested for model A in Ref.~\cite{Canet:2006xu}.} 
developed by Hohenberg and Halperin (for a review, see Ref.~\cite{Hohenberg:1977ym}).  
Within this approach we  examine the scale evolution of a stochastic equation of motion, 
which describes the critical dynamics of  slow modes.
These include fluctuations of the order parameter and of conserved quantities, 
which are relevant variables when addressing the long-wave length behavior of the system 
near the critical point. 
Since the transport coefficients are obtained from the corresponding response functions 
by taking the limit of both frequency and momentum to zero, 
they characterize the dynamics of the system in the low energy limit.

In analogy to the static case, 
the flow equations for transport coefficients derived from the DRG 
admit non-trivial fixed points, 
from which the {\it dynamical critical exponent}, $z$, 
and the dynamical scaling relations can be derived. 
The dynamical critical exponent, $z$, defines the characteristic frequency of the most relevant slow mode $\omega \sim k^z$, 
and the scaling relations link the singular contribution to the transport coefficients. 
From these properties one can deduce the singular behavior of the transport coefficients, in particular whether they diverge 
or remain finite at the critical point.
Based on the universal behavior, i.e. on the dynamical critical exponents and scaling laws, 
one identifies each system with a dynamical universality class. 
In contrast to the static case, 
the dynamical universality class is governed not only  
by the dimensionality, locality, and the symmetries of the system under consideration, 
but, in addition, by the properties of the relevant slow modes. 
Thus, the conservation or non-conservation of an order parameter, 
and the existence of mode-mode couplings among the slow modes affect the dynamical universality class. 
Therefore, even if two systems belong to the same static universality class, 
their dynamic universality class~\footnote{In what follows, we will frequently refer to the universality 
classes that were defined in Ref.~\cite{Hohenberg:1977ym}. Here we provide some properties of 
the relevant universality classes:
\begin{center}
\begin{tabular} {| c | c | c |}
\hline
Model & Slow mode(s) & Dynamical critical exponent in d=3 \\  
\hline
A & Non-conserved field  & $z_A = 2 + 0.7621 \eta'$     \\  
\hline
B & Conserved field & $z_B=4-\eta'$  \\  
\hline
C & N-component non-conserved field& $z_C=2+\alpha/\nu$ for $N=1$  \\  
  &  coupled to one component conserved field   &  $z_C=2+0.7621 \eta'$ for $N>1$  \\  
\hline
H & Conserved field coupled to & $z_H=4-18/19$  \\  
  & conserved transverse vector field &   \\  
\hline
G & N-component non-conserved field  & $z_G=3/2$  \\  
  & coupled to $N(N-1)/2$-component conserved field  &   \\  
\hline
\end{tabular}
\end{center}
Here $\alpha$ and $\nu$  are the static critical exponents and $\eta'$ is the anomalous dimension.
} may be different
\footnote{An example of such a situation is given by the models A and  B 
of critical dynamics
 (see Refs.~\cite{Hohenberg:1977ym,Folk:2006ve} for further details).
 In the static case,  both models, exhibiting  $Z(2)$  symmetry and belong to the static universality 
class of the Ising model in three dimensions. The dynamical universality class is, however, different. This difference, arises from 
the non-conservation (conservation) of the order parameter in model  A (B), and results in different long wavelength behavior characterized  by
the dynamical critical exponent:   $z_{\rm A}=2 + {\rm const}  \cdot \eta' $ 
and $z_{\rm B}=4 -\eta'$ where $\eta'$ is the anomalous dimension.
Another nontrivial example is the $O(N)$ model for $N=1$ (non-conserved order parameter) and 
model H  (conserved order parameter) of critical dynamics. 
These models also share the same static universality class, while the  dynamic 
universality classes differ. This implies a completely different  behavior of quantities such as the  shear viscosity, close to the critical point.
In model H, the shear viscosity diverges, while, as will be shown in this article, 
it is always finite in the $O(N)$ scalar field theory.
}. 
In this article, we determine the dynamical universality class of 
the $O(N)$ scalar field theory and show how the dynamical universality class depends on 
the number of components, $N$, and on the dimensionality, $d$. 

The paper is organized as follows: 
in the next section we identify the slow modes in the $O(N)$ scalar theory, 
and construct an effective Hamiltonian for them. 
In section ${\rm III}$ we review the static universality classification of the theory, 
and show that a non-trivial fixed point exists. 
In section ${\rm IV}$ 
we introduce the stochastic equation of motion, which 
describes the dynamics of the slow modes in the critical region.  
We then implement the DRG to find the fixed points
of the stochastic equations of motion , 
and determine the dynamical universality class. We close this section with a brief
discussion of the critical behavior of the bulk viscosity. 
Section ${\rm V}$ is devoted to summary, discussion, and outlook. Details on the 
derivation of the stochastic equation of motion and on the calculation of the response function are 
given in two appendices.

\section{Construction of the effective Hamiltonian}
Before considering the  dynamics of the theory, 
we have to build an effective Hamiltonian for the slow modes $A_l$. 
The probability distribution for the modes $A_l$ is given by the exponential of the Hamiltonian 
\footnote{Here, the prefactor in the exponent  $(k_B T)^{-1}$ is  
absorbed in the definition of the reduced effective Hamiltonian.}, $e^{-\mathcal H(\{A_l\})}$. 
The effective Hamiltonian defines the static critical behavior of the theory, 
and will  later on be incorporated in the equations of motion, 
from which we finally find the dynamical properties 
of the system close to the critical point. 
Although, the effective Hamiltonian, and the equations of motion for the slow modes 
have a microscopic origin, it is in general a very challenging problem to derive them  
starting from the microscopic Lagrangian. Therefore, in the present work, 
we formulate the effective Hamiltonian and the equations of motion on 
a phenomenological basis. The guiding principles in such a formulation 
are similar to those of Ginzburg-Landau theory. 
Note that, in our case, the slow variables in the Ginzburg-Landau Hamiltonian  
are all {\it fluctuations}, i.e., deviations of variables from their equilibrium values.

Candidates for the slow mode of the theory are  the 
fluctuations of the order parameter $\phi_i$, the
energy-momentum density, $E$ and $\vec{J}$, 
and the $O(N)$ charge density, $Q_{ij}$. 
Owing to the symmetry $Q_{ij}=-Q_{ji}$, there are $N(N-1)/2$ charges associated
with generators of the $O(N)$ group. 
The order parameter of the  theory is not conserved, while
the remaining variables (energy, momentum and $O(N)$ charge) 
are conserved quantities.

In the present work, we consider a system approaching 
the critical point from the symmetric phase. 
In this case, it is straightforward to construct the effective Hamiltonian for the slow modes 
\begin{eqnarray}
\label{H}
{\mathcal H}&=&
\int {\rm d}^dx \ldk   {\mathcal H}_{\phi}  
+\gamma_0  \phi_i^2 E  +\frac{1}{2} C_0^{-1}E^2 +\frac{1}{2} \vec{J}^2  +\frac{1}{2} \chi_0^{-1} Q_{ij}^2  +{\mathcal H}_{s}\rdk, \\
{\mathcal H}_{\phi} &=&  \frac{1}{2} \lk  \vec{\nabla}\phi_i \rk^2 
+\frac{r_0}{2} \phi_i^2 +\frac{u_0}{4} \lk \phi_i^2\rk^2,  
\label{Heff}\\ 
{\mathcal H}_{s}
&=&-\phi_i h_i-\vec{J}\cdot \vec{H} +\beta E -\mu_{ij} Q_{ij}, 
\end{eqnarray}
where ${\mathcal H}_{s}$ is the source term, 
which is introduced for later convenience.  
We follow the convention that repeated indices imply summations, e.g.,  
$Q_{ij}^2 \equiv \frac{1}{2} \sum_{ij=1}^N Q_{ij} Q_{ij}$. 
The effective Hamiltonian (\ref{H}) includes all possible candidates for slow modes in an $O(N)$ scalar field theory. 

Since the Hamiltonian includes up to quadratic terms in $E$, $\vec{J}$ and $Q_{ij}$, 
the original Hamiltonian density 
for the order parameter fluctuation, ${\mathcal H}_\phi$, 
is recovered after integrating out these variables and performing a suitable redefinition of the couplings. 
This implies that the critical statics of the Hamiltonian ${\mathcal H}$ is the same as that of ${\mathcal H}_\phi$. 

The coefficients of the  Hamiltonian~(\ref{H}), 
are given by the static susceptibilities of the slow modes. 
Since the susceptibility of the momentum current $\vec{J}$ always remains finite, 
we have absorbed the coefficient of $\vec{J}^2$ by a redefinition of the field $\vec{J}$. 
The $O(N)$ charge susceptibility, $\chi_0$, also remains finite for zero net charge
(i.e. zero chemical potential). 
In the case of Bose-Einstein condensation with a finite $O(N)$ charge, however, 
$\chi_0$ diverges at the critical point. 
We do not consider this situation, 
but keep $\chi_0$ explicitly in the Hamiltonian for later convenience. 

There are two contributions to the fluctuations of the energy density, 
$\delta E = T \delta S + h_i \delta \phi_i$, 
where $S$ is the entropy density\footnote{
In the remainder of this section, we explicitly denote the fluctuation of a variable $X$ by $\delta X$, in order to avoid ambiguities.}.
Consequently, the static correlation with the order parameter fluctuation is given by 
\beq
\label{Ephi}
\langle \delta E \delta \phi_i \rangle 
= T \langle \delta S \delta \phi_j\rangle 
+ h_j \langle \delta \phi_i \delta \phi_j \rangle. 
\eeq
In an $O(N)$ symmetric system (no explicit symmetry breaking), $h_j =0$ at the physical point. Therefore, 
the last term in Eq.~(\ref{Ephi}) does not contribute 
to $\langle \delta E \delta \phi_i \rangle$.  
The correlation $\langle  \delta S \delta \phi_i\rangle$ 
is nonzero at temperatures below $T_c$, 
where the symmetry is spontaneously broken  
in a specific direction of the field. Thus, for this component of the field $\phi_i$: 
$\langle \delta S \delta \phi_i \rangle 
\sim -\partial_T \langle \phi_i \rangle \neq 0$. 
Moreover, this quantity diverges close to the critical point in the broken phase, since
$\langle \delta S \delta \phi_i \rangle\sim t^{\beta-1}$, where $t=(T-T_c)/T_c$ is  the reduced  temperature,  
and the critical exponent $\beta \le 1/2$. 
However, at temperatures above $T_c$,
the correlation function $\langle  \delta S \delta \phi_i\rangle$ 
vanishes due to symmetry in the absence of the external field $h_i$. Indeed, since for $t> 0$ and $h_{i}\to 0$  
the order parameter scales like $\langle \phi_i \rangle \sim t^{-\gamma} h_i$, which implies that   
$\langle  \delta S \delta \phi_i\rangle\sim t^{-\gamma-1} h_i$ 
Thus, in the effective Hamiltonian, 
there is no bilinear contribution of the form $\sim \phi_i E$ for $t>0$. 

Now consider the autocorrelation function of the energy fluctuations
\beq
\langle \delta E \delta E \rangle =T^2\partial_T E,
\eeq
which is proportional to the specific heat, $C$.
Near the critical point, the singular part of $C$ scales  
as $\sim t^{-\alpha}\sim \xi^{\alpha/\nu}$ 
where $\xi$ is the correlation length.  
The specific heat is related to the static susceptibility of the energy, 
$C \propto  \chi_E(|\vk|=0;T)$,  up to some dimensionful factor.
The sign and numerical value of the critical exponent $\alpha$ depends on 
the number of field components, $N$, and the dimensionality, $d$ 
(see  e.g.~\cite{Fisher:1974uq}). 

In the effective Hamiltonian, we have dropped the spatial derivative terms, 
i.e. terms of the form $\lk \nabla_m {A_l}\rk^2$, for all fields $A_{l}$ 
which turn out to be irrelevant for long wavelength physics, except for the order parameter.
Consider for instance the term involving derivatives of the energy density, i.e. $\lk\vec{\nabla}E
\rk^2$. For negative $\alpha$, the specific heat remains finite at the critical point.  
Hence, the coefficient of the $E^{2}$ term in the Hamiltonian
scales as $C^{-1} \sim \xi^0$. Using standard renormalization group arguments,
one then finds that  the derivative term $\lk\vec{\nabla}E\rk^2$ is irrelevant. 
Also for positive $\alpha$, 
when the specific heat diverges as $\xi^{\alpha/\nu}$, the term is irrelevant as long as 
$\alpha/\nu < 2$.  This inequality is in general satisfied, since $\alpha/\nu$ is small,  $\mathcal{O}(\varepsilon)$, where $\varepsilon=4-d$. 
In the  case of interest, where the critical point is approached from the symmetric phase, i.e.  $T\to  (T_c)^{+}$ for $h=0$, 
all derivative terms of the conserved quantities are, by the same reasoning, negligible. 
Consequently, for static properties, contributions at the scale $\sim \Lambda$ are due
only to loop corrections involving fluctuations of the order parameter. The corresponding derivative term is relevant, yielding nontrivial contributions to the critical exponents through the nonzero anomalous dimension.

\section{Critical statics}

\subsection{Critical exponents and scaling hypothesis}
In this section we review the critical statics at continuous/second-order phase transitions \cite{Wilson:1973jj,Fisher:1974uq}. 
A general effective theory for the order parameter of a continuous phase transition 
was developed by Landau. This theory provides a mean-field description of the phase transition. 
The Ginzburg criterion defines the region of applicability of the mean-field approximation. 
Close to the critical point, in the critical region, the Ginzburg criterion is violated and mean-field theory 
breaks down. As the critical temperature is approached, low-energy fluctuations of the order parameter diverge owing to the flatness of the potential. Consequently, naive perturbation theory for loop corrections fails. 
One finds by dimensional analysis in terms of the correlation length $\xi$, 
that higher order interaction terms, e.g., the 4-point coupling, 
diverge as the critical point is approached for $d<4$, in particular in three dimensions.  
Therefore, a systematic analysis of the loop contributions in the critical region is in general 
difficult. 

In spite of these complications, various scaling relations have been found among the critical exponents. 
These relations imply that there are only a few independent critical exponents. 
In the $O(N)$ theory, there are two independent exponents 
associated with the reduced temperature and the external field. 
Except for the hyper scaling relations, the scaling relations hold for empirically determined exponents 
in critical region, but also for the Landau mean-field theory. 
Scaling relations are easily derived, 
once the general {\it assumption of homogeneity} is made
for the singular part of the thermodynamic potential density: 
\beq
F_s(t,h)=L^{-d} F_s(L^{\Delta_t} t, L^{\Delta_h} h),
\label{homo}
\eeq
where $L$ is an arbitrary number not much greater than unity, 
$d$ is the number of spatial dimensions, 
and $\Delta_{t, h}$ is the scaling dimension of the reduced temperature $t$ and the external field, 
$h$.
This hypothesis was established more rigorously by Kadanoff 
using block spin transformations for the Ising model. 
Later on Wilson developed a systematic method, applicable to any system, for evaluating scaling 
dimensions $\Delta_{t, h}$ explicitly. 
The latter is known as the renormalization group (RG) method 
with the epsilon expansion about the critical dimension~\cite{Wilson:1973jj,Fisher:1974uq}. 

The RG method consists of two steps: 
i) integrating out a high momentum shell $\Lambda/b < k < \Lambda$ with the parameter $b>1$, 
and ii) rescaling the unit length $k \rightarrow bk$ and other variables accordingly. 
Consecutive implementation of these procedures yields a
flow of the renormalized Hamiltonian (thermodynamic potential), 
i.e., a flow under the RG transformation in the full  parameter space ${V}$. 

A critical point of a continuous phase transition corresponds to a fixed point 
of the flow, where the  length scale $\xi$ goes to infinity. 
Let ${V^*}$ be a fixed point, and ${v}={V-V^*}$ the deviation from it. 
Then the thermodynamic potential density can be written as 
\beq
F({V})=F(V^*;\{v\}). 
\eeq 
Consider a system at a point in the parameter space, which is not a fixed point. 
After a single renormalization step, we obtain 
\beq
F(V^*;\{v_r\},\{v_{ir}\})\rightarrow 
b^{-d}F(V^*;\{b^{\Delta_{vr}}v_r\},\{b^{\Delta_{vir}}v_{ir}\}), 
\eeq
where the deviations $\{v\}$ can be classified into {\it relevant} $\{v_r\}$ and {\it irrelevant} parameters $\{v_{ir}\}$ 
according to their scaling dimension. 
By definition, the relevant (irrelevant) parameters have positive (negative) scaling dimension $\Delta_{vr}>0$ ($\Delta_{vir}<0$), 
and ones with vanishing scaling dimension are called {\it marginal} parameters.
The factor $b^{-d}$ in front of $F$ stems from the rescaling and reflects the dimensionality of $F$. 
Repeating this procedure $n$ times results in the substitution $b \rightarrow b^n$. 

Note that in general, a constant term appears in $F$ after the renormalization procedures. This term, 
which breaks homogeneity, originates from integrating out the higher momentum shells. 
However, since this term is non-singular, it can be dropped.
The remainder obeys the homogeneous relation: 
$
F_s(V^*;\{v_r\},\{v_{ir}\})\simeq
b^{-d}F_s(V^*;\{b^{\Delta_{vr}}v_r\},\{b^{\Delta_{vir}}v_{ir}\})
$.
The homogeneous scaling relation (\ref{homo}) holds 
only close to the critical point, in the so called scaling region.
Here the irrelevant variables are very small, and can be put to zero, since $b^{n \Delta_{vir}} \ll 1$. 
The singular behavior near the critical point is controlled only by the relevant parameters, 
and various scaling relations are obtained naturally, 
provided the system is sufficiently close to the critical point.
The relevant variables $\{v_r\}$ again, can be identified with the temperature and the magnetic field,  
$v_{r1}\propto t$ and $v_{r2}\propto h$.

\subsection{Static critical phenomena}
We first discuss renormalization of the effective potential $\Omega=-(\ln Z)/V$ per volume 
to define the static properties of the $O(N)$ scalar field theory in the low energy limit. 
The partition function is defined by $Z=\sum e^{-{\mathcal H}}$ 
with the dimensionless reduced Hamiltonian ${\mathcal H}$. 
The static renormalization group aims at tracing the evolution 
of the coefficients in the Hamiltonian (\ref{Heff}), 
under the RG transformation. 

The theory is defined with a finite ultraviolet cutoff $\Lambda$.  
This means that the $O(N)$ scalar field theory is an effective one, 
which can be applied only at scales below $\Lambda$. 
Since the soft modes are treated explicitly, the theory possesses the correct infrared behavior. 
We follow the renormalization group procedure developed by Wilson and Kogut~\cite{Wilson:1973jj}. 
This involves the two steps mentioned above:  
integration over the momentum shell $\Lambda/b \le k \le \Lambda$ 
in loops corrections with a parameter $b>1$, 
and  rescaling  the variables and fields
\beq
x &\rightarrow& x/b, \\ 
\Lambda &\rightarrow& b\Lambda, \\
\phi_i &\rightarrow& b^{a_\phi} \phi_i .
\eeq  
The scaling dimension of the order parameter field, $a_\phi$, 
is determined by the requirement that the rescaling leaves the auto-correlation function 
of $\phi_i$ unchanged, i.e., keeping the derivative term of $\phi_i$ to be marginal: 
$a_\phi=\frac{1}{2}\lk d-2+\eta'\rk$.  
Here $\eta'$ is the anomalous dimension, not be confused with the shear viscosity, $\eta$.

This procedure provides an 
evolution of the system 
under successive changes of the length scale and decimation of shorter wavelength modes. 
This process generates all couplings including higher order ones allowed by the symmetry of the system.
The theory approaches a low-energy effective theory for the long wavelength modes.

After repeating the renormalization procedure  $l$ times, 
one obtains the well-known recursion relations for the coefficients, 
to leading order in the coupling $u$, 
\beq
r_{l+1}
&=& 
b^{d-2 a_\phi} \ldk r_l + 2(N+2) \Omega_4 u_l 
\ltk \Lambda^2 \lk 1-b^{-2}\rk -2 r_l \ln b \rtk \rdk \\ 
u_{l+1}
&=&
b^{d-4 a_\phi} u_l \ldk  1-4(N+8)\Omega_4 u_l \ln b\rdk. 
\eeq
The above relations are obtained 
for $d=4-\varepsilon$ dimensions. 
The factor $\Omega_d=2^{1-d} \pi^{-d/2} \Gamma(d/2)$  originates from 
the solid angle integration in $d$ dimensions, divided by $(2\pi)^d$, 
with $\Gamma(x)$ being the Gamma function. 
In the right hand side of the recursion  relations, the factors of $b$ with exponents stem from the rescaling, 
while the terms proportional to $\Omega_4$ arise in the decimation of shorter wavelength modes.
These two contributions play a competitive role in the  RG evolution. This makes  an appearance 
of non-trivial fixed points possible.
A simple dimensional analysis shows that interaction terms higher than quartic are irrelevant 
under the renormalization. 
The recursion relations in fact admit a non-trivial critical fixed point, 
\beq
r^* &=& -\frac{1}{2} \varepsilon \frac{N+2}{N+8} \Lambda^2 + O(\varepsilon^2), \\
u^* &=& \frac{\varepsilon}{4\Omega_4(N+8)} +O(\varepsilon^2), 
\eeq
implying that the system undergoes a second-order phase transition 
with infinite correlation length $\xi$. 

One can extract the scaling dimensions by observing how the coupling parameters behave near the fixed point. 
To do this, it is sufficient to linearize the recursion relations 
in terms of $\delta r_l\equiv \lk r^*-r_l\rk/\Omega_4\Lambda^2$ and $\delta u_l\equiv u^*-u_l$: 
\beq
\lk \begin{array}{cc} 
\delta r_{l+1}-\delta r_l\\
\delta u_{l+1}-\delta u_l
\end{array}\rk \simeq \ln b 
\lk \begin{array}{cc} 
2-\frac{N+2}{N+8}\varepsilon & \ 4(N+2)\ldk 1+\frac{N+2}{2(N+8)}\varepsilon\rdk\\
0 & -\varepsilon 
\end{array}\rk 
\lk \begin{array}{cc} 
\delta r_l\\
\delta u_l
\end{array}\rk. 
\eeq
Then eigenvalue problem of the above matrix tells 
that only $r$ is the relevant parameter and $\delta r \propto b^{\Delta_r}$ 
with $\Delta_r=2-\frac{N+2}{N+8}\varepsilon +O(\epsilon^2)$ being the scaling dimension, 
while $u$ is irrelevant, with a negative scaling dimension $\Delta_u=-\varepsilon +O(\epsilon^2)$.

We also note that in the long-wave length limit, the self-interaction of the field $\phi_{i}$ vanishes in $d=4$ because 
$u^* \sim O(\varepsilon)$. 
Thus, the  perturbative expansion in the coupling constant $u$ is equivalent to an expansion in $\varepsilon$. This 
expansion is valid 
near the fixed point in a dimension slightly below four.  
In dimensions higher than four, the fluctuation contribution to the renormalization of 
4-point coupling is negligible, i.e. the mean-field description remains valid. 
The physical correspondence of dimensionful quantity $r-r^*$ with thermodynamic variables 
is introduced by hand, 
e.g.,  $r-r^*\propto T-T_c$ near the critical point.  

Let us now examine the interaction term $\gamma_0 \phi_i^2 E$. 
Since only this term provides the static coupling 
between order parameter and energy density fluctuations, 
its critical behavior is crucial in the subsequent analyses of critical dynamics. 
A system with a non-conserved order parameter coupled to the conserved energy was classified 
by Hohenberg and Halperin (see Ref.~\cite{Hohenberg:1977ym} and references therein), as model C. 
The recursion relations for $\gamma_0$ and $C_0$ are given by 
\beq
C_{l+1}^{-1}&=&b^{d-2a_E} C_l^{-1}\ldk 1-2 N v_1 \ln b \rdk, \\
v_{l+1}&=&b^{d-4a_\phi} v_l\ldk 1-8(N+2) \Omega_4 u_l \ln b -2 N v_l \ln b \rdk,  
\eeq
where $v_0 \equiv \Omega_4 \gamma_0^2 C_0$ is the dimensionless three-point coupling, 
and $d-2 a_E = \tilde{\alpha} /\nu  = \alpha \theta(\alpha)/\nu$ 
with $\alpha$ being the exponent of specific heat $C\sim \lk T-T_c\rk^{-\alpha}\sim t^{-\alpha}$.
Here $\theta(x)$ is the unit step function and 
$\alpha$ is the critical exponent of the specific heat. 
The fixed point of the coupling $v$ is given by 
$v^*= \tilde{\alpha}  +O(\varepsilon^2)$, which vanishes for negative $\alpha$. 
The sign and value of the critical exponent $\alpha$ depends on $N$ and $d$, 
as noted above. We return to this point in the subsequent section.

For $\alpha>0$, fluctuations of the energy can become critical, 
i.e. the corresponding mass (the inverse of the specific heat) vanishes at the critical point. 
Thus, also the critical dynamics may be affected by energy fluctuations. 
On the other hand, for $\alpha<0$ the mass term remains finite and fluctuations of the  
energy do not affect the static critical properties of other variables. 
Nevertheless, since the order parameter always exhibits critical fluctuations at a second-order transition, 
it is possible that these fluctuations affect other variables 
through dynamical effects, like mode-mode couplings.

\section{Critical dynamics}

\subsection{The stochastic equation of motion}
To address the critical dynamics of a system, one needs the equations of motion \cite{zwa1,Mori1,Mori2}.  
The low energy and long-wave length dynamics in the critical region 
is dominated by slow modes, i.e. 
fluctuations of the order parameter and the conserved quantities. 
We describe such modes by fields $A_l(t, \vec{x})$ varying in space and time,  
and introduce a stochastic equation of motion to describe the dynamics of the fields.
In the mixed Fourier representation 
\beq
\partial_t A_l(t,\vk) = L_{l m}(\vk) 
\frac{\delta {\mathcal H}}{\delta A_m(t,\vk)} 
-\ldk A_l, A_m \rdk_{PB} \frac{ \delta {\mathcal H}}{\delta A_m(t,\vk)} 
+ \Theta_l(t,\vk). 
\label{SEM}
\eeq
Here ${\mathcal H}={\mathcal H}\lk \ltk A_l\rtk \rk$ 
is a reduced effective Hamiltonian, 
which is a functional of the slow modes, 
and $e^{-{\mathcal H}}$ is proportional to the probability for a particular configuration  
of the fields $A_l$. 
The first term on the right side involves transport coefficients $L_{l m}(\vk)$, which 
are responsible for the damping of fluctuations. 
Hence, this term describes irreversible processes.  Owing to this term and the noise term $\Theta_l$, 
the system eventually reaches an equilibrium state where 
$\delta {\mathcal H}/\delta A_l=0$. The noise term satisfies  the fluctuation-dissipation relation, 
\beq
\langle \Theta_l(t,\vk) \Theta_m(t',\vk') \rangle 
&=& 2 L_{l m}(\vk) \delta(t-t')\delta(\vk-\vk'), 
\label{fluct-diss}
\eeq
which is valid for Gaussian noise. 
The cross terms with $l \neq m$ originate from 
a possible bilinear mixing among the variables, $\sim A_l A_m$, 
in the effective Hamiltonian ${\mathcal H}$. 

The second term, with the Poisson bracket, $\ldk \cdots \rdk_{PB}$, 
yields non-linear interactions, 
the mode-mode couplings \cite{Fix,Kawa}. 
These describe the non-dissipative (reversible) processes, which are
responsible for the large amplitude collective fluctuations induced 
by the critical behavior of the order parameter. 
Consequently, this term contributes to the singularities, which define the critical dynamics. 
The mode-mode couplings are formulated in terms of the generators of the relevant symmetries, and 
thus preserve the invariances of the original equations of motions.

The equation of motion can be derived from the Liouville equation 
by  the projection method in the Markovian approximation 
under some reasonable  assumptions. 
The derivation is reviewed in appendix A. Further  details  can be found in Ref.~\cite{zwa}.  
The presence of the Poisson bracket implies 
that the equations were derived from Hamilton's equations of the classical theory, 
which is valid for slow modes (see also discussion in Ref.~\cite{Berges:2009jz}).

\subsection{Response functions and transport coefficients}

A set of stochastic equations of motion for the slow modes $A_l = \{\phi_i, E, \vec{J}, Q_{ij}\}$ is obtained 
from Eq.~(\ref{SEM}), given the effective Hamiltonian constructed above (\ref{Heff}), 
\beq
\frac{\partial \phi_i}{\partial t} 
&=& 
-\lambda_0 \frac{\delta {\mathcal H}}{\delta \phi_i} 
- g_0 \vec{\nabla} \phi_i 
\cdot \frac{\delta {\mathcal H}}{\delta \vec{J}} 
+\tilde{g}_0 \ldk \phi_i, Q_{jk}\rdk_{PB} 
\frac{\delta {\mathcal H}}{\delta Q_{jk}}+ \theta_i,  \\
\frac{\partial E}{\partial t} 
&=& 
\Gamma_0 \vec{\nabla}^2\frac{\delta {\mathcal H}}{\delta E} 
-g_0 \vec{\nabla} E 
\cdot \frac{\delta {\mathcal H}}{\delta \vec{J}} + \theta_E,  \\
\label{jEoM}
\frac{\partial \vec{J}}{\partial t} 
&=& \mathcal{T}\cdot \ldk 
\eta_0 \vec{\nabla}^2 \frac{\delta {\mathcal H}}{\delta \vec{J}} 
+ g_0 \vec{\nabla} \phi_i 
\frac{\delta {\mathcal H}}{\delta \phi_i} 
+g_0  \vec{\nabla} E 
\frac{\delta {\mathcal H}}{\delta E}
+g_0 \vec{\nabla} Q_{ij} 
\frac{\delta {\mathcal H}}{\delta Q_{ij} }+ \vec{\theta}_J\rdk,   \\
\frac{\partial Q_{ij}}{\partial t} 
&=& 
\Pi_0 \vec{\nabla}^2\frac{\delta {\mathcal H}}{\delta Q_{ij}} 
+\tilde{g}_0 
\ldk Q_{ij}, Q_{kl} \rdk_{PB} 
\frac{\delta {\mathcal H}}{\delta Q_{lk}}  
+\tilde{g}_0 \ldk Q_{ij}, \phi_k \rdk_{PB} 
\frac{\delta {\mathcal H}}{\delta \phi_k} \nn 
&&- g_0 \vec{\nabla} Q_{ij} 
 \cdot \frac{\delta {\mathcal H}}{\delta \vec{J}}+ \theta_{ij}, 
\eeq
where $g_0$ and $\tilde{g}_0$ are the mode-mode couplings associated with the translation and $O(N)$ symmetries and 
$\mathcal{T}={\bf 1}-\frac{\vec{\nabla}\vec{\nabla}}{\vec{\nabla}^2}$ is 
the projection operator on the transverse direction. 
Fluctuations of the transverse momentum describe diffusive modes, 
while fluctuations of the longitudinal momentum coupled 
with energy fluctuation describe sound waves, which have a linear dispersion relation. 
The latter are not taken into account  
because the sound mode corresponds to fast dynamics, 
which does not affect late time evolution. 
The longitudinal momentum, however, has to be considered  in order to 
address the critical behavior of the bulk viscosity. 
This will be discussed in section IV. 

The Poisson bracket between  $Q_{ij}$ and $\phi_i$ are deduced 
from the quantum commutation relations 
\beq
\ldk \phi_i, Q_{jk} \rdk_{PB}
&=&-\phi_j\delta_{ki} +\phi_k \delta_{ij}, \\
\ldk  Q_{ij}, Q_{kl} \rdk_{PB}
&=& 
-Q_{jl} \delta_{ik}+Q_{jk} \delta_{il}
+Q_{il} \delta_{jk}-Q_{ik} \delta_{jl}. 
\eeq
In line with (\ref{fluct-diss}), the noise term correlation functions satisfy 
\beq
&&\langle \theta_i(t,x) \theta_j(t',x') \rangle
=2\lambda_0 \delta_{ij} \delta^d(x-x') \delta(t-t'),\\
&&\langle \theta_E(t,x) \theta_E(t',x') \rangle
=-2\Gamma_0 \vec{\nabla}^2 \delta^d(x-x') \delta(t-t'),\\
&&\langle \vec{\theta}_{J}(t,x) \vec{\theta}_{J}(t',x') \rangle
=-2\eta_0 {\bf 1}\vec{\nabla}^2 \delta^d(x-x') \delta(t-t'), \\
&&\langle \theta_{ij}(t,x)  \theta_{kl}(t',x') \rangle
=-2\Pi_0 \lk \delta_{ik}\delta_{jl}-\delta_{il}\delta_{jk}\rk 
\vec{\nabla}^2 \delta^d(x-x') \delta(t-t'). 
\eeq

The transport coefficients are obtained 
from low energy limit of the dynamical response functions. 
In frequency and momentum space, 
they are given by
\beq
\frac{1}{\lambda}&=&
i \lim_{k \rightarrow 0} \frac{\partial \chi_\phi^{-1}(k)}{\partial\omega}, 
\label{relam}\\
\frac{1}{\Gamma}&=&
i \lim_{k \rightarrow 0} \vk^2 \frac{\partial \chi_E^{-1}(k)}{\partial\omega}, 
\label{relam1}\\
\frac{1}{\eta}&=&
i \lim_{k \rightarrow 0} \vk^2 \frac{\partial \chi_J^{-1}(k)}{\partial\omega}, 
\label{reeta}\\
\frac{1}{\Pi}&=&
i \lim_{k \rightarrow 0} \vk^2 \frac{\partial \chi_Q^{-1}(k)}{\partial\omega}, 
\label{repi} 
\eeq
where we use the short hand notation $k\equiv \{ \omega, \vk \}$.
The limit is taken first with respect to frequency 
and then to  momentum. 
The response functions are obtained from the solution 
of the stochastic equations of motion, after averaging over the noise $\left\langle \cdots \right\rangle_\theta$, 
\beq
\chi_\phi(k)_{ij}
&=&
\left\langle \frac{\delta \phi_i(k)}{\delta h_i(k)} \right\rangle_\theta \delta_{ij},\\
\chi_E(k)
&=&
-\left\langle \frac{\delta E(k)}{\delta \beta(k)}\right\rangle_\theta, \\
\chi_J(k) {\mathcal T}_{ij}
&=&
\left\langle \frac{\delta \vec{J}_i(k)}{\delta \vec{H}_j(k)}\right\rangle_\theta, \\
\chi_{Q_{ij}}(k) 
&=&
\lk \delta_{ik}\delta_{jl}-\delta_{il}\delta_{jk}\rk 
\left\langle \frac{\delta Q_{ij}(k)}{\delta \mu_{kl}(k)}\right\rangle_\theta. 
\eeq 
The source terms are put to zero, after the variations in the above equations.

\subsection{Dynamical renormalization group and dynamic scaling}
In this section we investigate the fixed points of the stochastic equations of motion 
for the $O(N)$ scalar field theory, using the dynamical renormalization group (DRG)~\footnote{
In analogy to the static case, a fixed point of the equations of motion is necessary to be able 
to sort out the critical dynamics.}. 
We thus determine the universal properties of the critical dynamics. 
The procedure of the DRG is very similar to the static one: 
i) in loop corrections, the momentum shell $\Lambda/b \le |\vk| \le \Lambda$ is integrated out, 
while the frequency is integrated over the whole domain of definition 
$-\infty \le k_0 \le \infty$, 
and ii) rescaling of all variables. 
A difference from the static renormalization is that there appears a frequency scale,  
and its scaling in length units is assumed to be $\omega \rightarrow b^z \omega$ 
where $z$ is the dynamical critical exponent for the slowest mode. 
The rescaling factors in the $O(N)$ scalar field theory follow from 
dimensional analysis of the stochastic equations of motion and 
of the effective Hamiltonian,  ${\mathcal H}$,  which is a dimensionless quantity, 
\beq
x &\rightarrow& b^{-1}x, \\
\Lambda &\rightarrow& b\Lambda, \\
t &\rightarrow& b^{-z}t, \\
\phi &\rightarrow& b^{a_\phi}\phi, \\
E &\rightarrow& b^{a_E}E, \\
\vec{J} &\rightarrow& b^{a_J}\vec{J}, \\
Q &\rightarrow& b^{a_Q}Q.
\eeq
The exponents for the fields are 
$a_E=(d-\tilde{\alpha}/\nu)/2$, $a_J=d/2$, $a_Q=d/2$, 
and as in the static case $a_\phi=(d-2+\eta')/2$. 
We have set the scaling dimension of $\chi_0$ to zero. 

In evaluating the dynamical response function, we employ a loop expansion 
in terms of the deviation from the upper critical dimension $\varepsilon=4-d$ 
in the same way as in the static case. 
For instance, the response function of the order parameter is expressed as
\beq
\chi_\phi(k)
&=&
\left \langle \left. \frac{\delta \phi(k)}{\delta h(k)}\right|_{h\rightarrow 0}
\right\rangle_{\theta}
=
G_\phi(k) \ldk  \lambda +\Sigma_\phi(k)\rdk \label{chiphi}, 
\eeq
where $\Sigma$ represents the loop corrections integrated over the momentum shell and the bare propagator is given by
\beq
 G_\phi(k)
 &=&\frac{1}{-i\omega +\lambda_0 \lk r_0+\vec{k}^2\rk}. 
\eeq
A renormalized relaxation rate for the order parameter fluctuation, $\lambda$, 
is derived from the response function $\chi_\phi$ (\ref{chiphi}). 
This procedure corresponds to a single renormalization operation.
Thus, the recursion relation for $\lambda$ reads 
\beq
\lambda^{-1}_{l+1}=b^{2-z-\eta'}\lambda^{-1}_{l} \left[ 1+ \tilde{\Sigma}_\phi(\lambda_l,\Gamma_l,\cdots;b) \right],
\eeq
where $\tilde{\Sigma}$ is a dimensionless loop function, 
and the overall rescaling factor in $b$ can be determined from the rescaling factors of the other variables 
using the equation of motion. 
The recursion relations for the remaining transport coefficients follow the same procedure. 
See the following sections and Appendix B for details.

\section{Results for transport coefficients}

\subsection{Flow equation}

In the dynamical renormalization procedure presented in the previous sections, 
we derived  a set of recursion relations for transport coefficients to one loop order 
\beq
\lambda_{l+1}
&=&b^{z-2+\eta'}\lambda_l \ldk 1 
-\frac{4\gamma_l^2 C_l \lambda_l}{\lambda_l+\Gamma_l/C_l}\Omega_4 \ln b
+\frac{h_l^2 \lk N-1\rk}
{\lambda_l \chi_l\lk \lambda_l+\Pi_l/\chi_l\rk}\Omega_4 \ln b \rdk, \\\
\Gamma_{l+1}
&=& 
b^{z-2-\bar{\alpha}/\nu}\Gamma_l 
\ldk 1 
+\frac{3}{4}\frac{g_l^2}{\Gamma_l C_l^{-1}\lk \Gamma_l C_l^{-1}+\eta_l \rk}
\Omega_4\frac{\Lambda^2}{2}\lk 1-b^{-2}\rk \rdk, \\
\eta_{l+1}
&=&b^{z-2}\eta_l\ldk 1+ 
\frac{g_l^2}{24\lambda_l\eta_l} \Omega_4
\frac{\Lambda^2}{2}\lk 1-b^{-2}\rk\rdk, \\
\Pi_{l+1}
&=&
b^{z-2}\Pi_l\ldk 1+\frac{3g_l^2}{4\lk\Pi_l+\eta_l\rk\Pi_l}
\Omega_4\Lambda^2\lk 1-b^{-2}\rk
+\frac{\tilde{g}_l^2}{2\lambda_l\Pi_l}\Omega_4\ln b\rdk.
\eeq
The corresponding relations  for the mode-mode couplings and static coefficients read 
\beq
g_{l+1}&=&b^{z-3+\varepsilon/2}g_l, \\
\tilde{g}_{l+1}&=&b^{z-2+\varepsilon/2}\tilde{g}_l, \\
C_{l+1}^{-1}
&=&b^{d-2a_E} C_l^{-1} \ldk 1-2NC_l \gamma_l^2 \Omega_4 \ln b \rdk, \\
\gamma_{l+1}
&=&b^{d-2a-a_E} \gamma_l \ldk 1-4\lk N+2\rk u_l^2 \Omega_4 \ln b -2N\gamma_l^2 C_l\rdk, \\
\chi_{l+1}&=&\chi_{l},
\eeq
where $a_\phi$ and $a_E$ were defined above. 
Note, that the mode-mode couplings $g$ and $\tilde{g}$ exhibit only trivial scaling
without loop corrections. This follows from   
Ward identities for the higher order response functions. 
This  can be  also understood from Galilean invariance and invariance of the equations of motion 
under $O(N)$ rotations. 
Using these recursion relations, we find the fixed points of the equations of motion, 
and extract dynamical critical exponents and scaling relations. 

In the continuum limit, $b\rightarrow 1$, the recursion relations yield  
the flow equations,
\beq
\partial \lambda&=&\lambda \left[ z-2+\eta'
-\frac{4v^*}{1+\omega_1}+(N-1)\frac{f_2}{1+\omega_2}\right],\\
\partial \Gamma&=&\Gamma \left[ z-2-\frac{\bar{\alpha}}{\nu}
+\frac{3}{4}\frac{f_3}{1+\omega_3}\right], \\
\partial \eta&=&\eta \left[ z-2+\frac{1}{24}f_1 \right],\\
\partial \Pi &=& \Pi \left[ z-2 +\frac{3}{2} \frac{f_4}{1+\omega_4}+\frac{1}{2} f_2 \right], 
\eeq
where $\partial=\frac{\partial}{\partial \delta}$ with $b=1+\delta$. 
We introduce effective vertices for the mode-mode couplings: 
$f_1=\frac{g^2}{\eta\lambda} \Omega_4 \Lambda^2$, 
$f_2=\frac{\tilde{g}^2}{\lambda\Pi} \Omega_4$, 
$f_3=\frac{g^2}{\eta\Gamma C^{-1}} \Omega_4 \Lambda^2$, 
and 
$f_4=\frac{g^2}{\eta\Pi \chi^{-1}} \Omega_4 \Lambda^2$, 
\beq
\label{f1}
\partial f_1&=&f_1 \left[-2+\varepsilon-\eta'
+\frac{4v^*}{1+\omega_1} - (N-1)\frac{f_2}{1+\omega_2} -\frac{1}{24}f_1\right], \\
\label{f2}
\partial f_2&=&f_2 \left[\varepsilon-\eta'
+\frac{4v^*}{1+\omega_1} - (N-1)\frac{f_2}{1+\omega_2} -\frac{1}{2}f_2 
-\frac{3}{2}\frac{f_4}{1+\omega_4}\right], \\
\label{f3}
\partial f_3&=&f_3 \left[-2+\varepsilon+2Nv^*
-\frac{3}{4}\frac{f_3}{1+\omega_3}-\frac{1}{24}f_1\right], \\
\label{f4}
\partial f_4&=&f_4 \left[-2+\varepsilon 
-\frac{3}{2}\frac{f_4}{1+\omega_4}-\frac{1}{2}f_2-\frac{1}{24}f_1\right]. 
\eeq
and ratios of the transport coefficients, 
$\omega_1=\frac{\Gamma C^{-1}}{\lambda}$, 
$\omega_2=\frac{\lambda}{\Pi\chi^{-1}}$, 
$\omega_3=\frac{\Gamma C^{-1}}{\eta}$,
 and 
$\omega_4=\frac{\Pi \chi^{-1}}{\eta}$, 
\beq
\partial \omega_1&=&\omega_1 \left[ -\eta'-2Nv^*
+\frac{3}{4}\frac{f_3}{1+\omega_3}+\frac{4v^*}{1+\omega_1} 
-(N-1)\frac{f_2}{1+\omega_2}\right], \\
\partial \omega_2&=&\omega_2 \left[ \eta'-\frac{3}{2}\frac{f_4}{1+\omega_4}-\frac{1}{2}f_2 
-\frac{4v^*}{1+\omega_1} +(N-1)\frac{f_2}{1+\omega_2}\right], \\
\partial \omega_3&=&\omega_3 \left[ -2Nv^*
+\frac{3}{4}\frac{f_3}{1+\omega_3}-\frac{1}{24}f_1 \right], \\
\partial \omega_4&=&\omega_4 \left[\frac{3}{2}\frac{f_4}{1+\omega_4}+\frac{1}{2}f_2 
-\frac{1}{24}f_1 \right].
\eeq
Here the static fixed point of the three point function $v^*={\tilde{\alpha}} /( {2N\nu} )$ has been inserted. The three point vertex is of order $\varepsilon$, 
with $\alpha/\nu=(4-N)\varepsilon/(N+8)+O(\varepsilon^2)$ near four dimensions. 

The flow equations for the mode-mode couplings  
show that, except for $f_2$, 
there are contributions of order $O(\varepsilon^0)$ on the right hand side. 
Since these equations admit only trivial stable fixed points, i.e. $f_1^*=f_3^*=f_4^*=0$,  these mode-mode couplings vanish in the long wavelength limit.

In the classification of the dynamical universality class
we must, as implied by the discussion above, consider two cases, depending on the sign of $\alpha$. 
The critical number $N_c$, where $\alpha$ changes sign, is  given by Fischer \cite{Fisher:1974uq}: $\alpha$ is positive for $N<N_c$, with
$N_c \simeq 4 (1-\varepsilon)$ near four dimensions, and 
$N_c \simeq 1.8$ for $d=3$. 

\subsection{Fixed point for $N=1$}
We first consider the $N=1$ case, where the symmetry is reduced to the  discrete $Z_2$ symmetry, 
and the energy fluctuation must be taken into account, owing to the small but positive exponent $\alpha>0$.
The fixed points can be found by setting the right hand side of the flow equations to zero. To  leading order in $\varepsilon$ 
we find:
\beq
&&\lambda \left[  z-2
-\frac{2\alpha}{\nu}\frac{1}{1+\omega_1} \right]=0,\\
&&\Gamma \left[  z-2 -\frac{\alpha}{\nu} \right]=0, \\
&&  \frac{ \omega_1 \alpha}{\nu}  \left[ \frac{2}{1+\omega_1}-1  \right]=0,
\eeq
which admit a stable fixed point, 
\beq
&&\omega_1^*=1, \\
&&z=2+\frac{\alpha}{\nu}. 
\eeq
The last equation defines the dynamical critical exponent, which was deduced from the condition 
that $\Gamma$ and $\lambda$  each have a finite non-trivial fixed point. 
Thus, the long-wavelength dynamics of the system is, up to order $\varepsilon$, governed  
by fluctuations of the energy and the order parameter on equal footing. 
In the critical limit $\xi \rightarrow \infty$  (keeping $\xi k$ finite) we obtain
the following relaxation rates 
\beq
&&\delta \phi(t) \sim \exp(-\lambda \chi_\phi(k)^{-1} t) \sim \exp({-k^{2+\alpha/\nu} t}),\\
&&\delta E(t) \sim \exp({-\Gamma C^{-1} k^2 t}) \sim \exp({-k^{2+\alpha/\nu} t}),
\eeq
where we have used the fact that 
$\lambda \sim \xi^{2-z-\eta'}\sim \xi^{-\alpha/\nu+O(\varepsilon^2)}$, 
the order parameter susceptibility  $\chi_\phi(k)\sim k^{-2+\eta'}$, 
and $\Gamma \sim \xi^{2-z+\alpha/\nu}\sim \xi^{0+O(\varepsilon^2)}$. 
This result\footnote{The dependence of transport coefficients 
(and of other physical quantities) on the coherence length (temperature) 
are derived as follows: let $\{a\}$ be the full set of parameters (static coefficients) including relevant and irrelevant ones. 
Now we pick up only one relevant parameter $a_1$, e.g., the reduced temperature $a_1 \propto t$, 
and set the other relevant parameters on the critical surface, i.e., to zero. 
Since $\xi=\xi(\{a\})$, a transport coefficient 
$\Gamma=\Gamma(\{a\})=\Gamma\lk \xi, \{\bar{a}\}\rk$, 
where $\{\bar{a}\}$ represents the irrelevant parameters. We drop the irrelevant parameters
assuming that the system is very close to the critical point, 
and that the RG flow is sufficiently developed 
so the the parameters are in the immediate vicinity of the corresponding fixed point. 
Then, an RG transformation  changes $\xi \rightarrow \xi/b$ and $\Gamma \rightarrow b^X \Gamma$. 
One thus finds the scaling relation $b^X\Gamma(\{a\})=\Gamma\lk \xi/b, \{b^{\bar \Delta}\bar{a}\} \rk$ 
which leads to $\Gamma \sim \xi^{-X}$.}
can also be obtained  from the fixed point of $\omega_1^*$, 
which is the ratio of these two fluctuating modes, i.e.,  
$\omega_1\sim \Gamma C^{-1}/\lambda \sim \xi^{\eta'}$.  Thus, to leading order in $\varepsilon$, the 
critical exponent of $\lambda$ is smaller than that of $\Gamma$ by $\alpha/\nu$. 

It follows from the discussion above that, owing to the dominance of the fluctuations 
of the non-conserved order parameter and the conserved energy, 
the single component scalar theory belongs to the dynamic universality class of model C. 
In Ref.~\cite{Berges:2009jz}, 
the same conclusion was drawn based on the solution of a classical relativistic $\phi^4$ theory
on the lattice in $d=2$ spatial dimensions. 

An important point, which was not discussed so far 
is the renormalization flow of the shear viscosity. In the regime where $z=2+\alpha/\nu$, $\eta$ 
does not reach a finite stable fixed point. 
This means that the critical dynamics of the order-parameter does not affect the shear fluctuations.
Only short wavelength processes (rapid processes) contribute. 
Consequently,  the shear viscosity remains finite, 
in contrast to model H, where a finite fixed point of 
a mode-mode coupling provides the scaling relation between 
the exponents of heat conductivity and shear viscosity.

In order to obtain  a finite fixed point of the flow equation for $\eta^*$, $\partial \eta=\eta \ldk z-2 \rdk$, we have to set $z=2$, which is smaller than that of the fluctuating modes 
of the order parameter and the energy . 
Thus, long-wavelength fluctuation of the transverse momentum diffuses faster than the other modes, since 
$\delta J(t)\sim \exp({-\eta_0 \chi_\eta^{-1} k^2 t}) \sim \exp({-k^2 t})$, 
where we have used a bare shear viscosity $\eta_0 \sim \xi^0$
 and susceptibility $\chi_\eta \sim \xi^0$.  
Therefore, fluctuations of the transverse momentum correspond to a faster mode and decouple in the 
long-wavelength dynamics inside the critical region. 

\subsection{Fixed point for $N>1$}
For $N>1$
the static coupling between the energy density and the order parameter 
vanishes at the fixed point, as shown by Hohenberg and Halperin~\cite{Hohenberg:1977ym}
(more precisely the effective three-body coupling $\gamma_0^2 C_0$ vanishes)\footnote{
As explained earlier, the critical dynamics is governed by the sign of the critical exponent, $\alpha$. 
The absolute value of $\alpha$ is small for not too large N. Consequently, 
the sign of $\alpha$ is very sensitive to the approximations used. It is well known, that, to leading order, 
the epsilon expansion results in spurious sign of  $\alpha$ in the range $2<N<4$~\cite{Halperin:1974zz}. 
To obtain a physically correct result, we use the input from non-perturbative methods according 
to which $\alpha$ in $d=3$ is positive only for N=1, and negative otherwise.   
}.
 Therefore, the critical fluctuations of the order parameter do not 
directly affect the energy-momentum dynamics in the static case. In the dynamic case, such a coupling could be induced by 
the mode-mode coupling $f_1$, which, however, vanishes in the long-wavelength limit.  
Moreover, critical fluctuations of the order parameter couple 
to the $O(N)$ charge density only via the mode-mode coupling $f_2$. 
Thus, for $N>1$ one expects the energy modes to be irrelevant, while the $O(N)$ charge fluctuations affect the critical dynamics owing to the mode-mode coupling $f_{2}$.
Taking these arguments into account, we find the fixed points of the flow equations
 in the same way as for $N=1$ case: 
\beq
&&\lambda\left[ z-2+\frac{N-1}{1+\omega_2}f_2 \right]=0, \\
&&\Pi\left[ z-2+\frac{1}{2}f_2 \right]=0, \\
&&f_2\left[ \varepsilon -\frac{N-1}{1+\omega_2}f_2 -\frac{1}{2}f_2\right]=0, \\
&&\omega_2  f_2 \left[ \frac{N-1}{1+\omega_2} -\frac{1}{2} \right]=0. 
\eeq 
These equations yield  the following  stable fixed point and dynamical exponent, 
\beq
&&\omega_2^*=2(N-1)-1, \\
&&f_2^*=\varepsilon, \\
&&z=2-\frac{\varepsilon}{2}=\frac{d}{2}. 
\eeq
The dynamical exponent is obtained by requiring that 
$\lambda$ and $\Pi$ have a non-trivial fixed point.  
At the critical point, the transport coefficients scale  as 
$\lambda \sim \xi^{\varepsilon/2}$ and $\Pi\sim \xi^{\varepsilon/2}$ to leading order in $\varepsilon$. These results  
are consistent with the  fixed point of the mode-mode coupling: 
$f_2 \sim \xi^{-\varepsilon+\eta'}$. 
Long-wavelength fluctuations 
of the order parameter and of the $O(N)$ charge  
fall off with a characteristic frequency $\omega_k \sim k^{d/2}$. 

Fluctuations of energy and transverse momentum are governed by the 
flow equations $\partial \eta=\eta(z-2)$ and $\partial \Gamma=\Gamma(z-2)$. The fixed point at 
$z=2$ implies that these fluctuations are slower 
than those of the order-parameter and the $O(N)$ charge with $z=d/2<2$ 
in $d<4$ dimensions.  
However, from the fixed point analysis we see that the critical fluctuations of the 
order parameter do not affect the energy and transverse momentum fluctuations in the long wavelength limit. 
Thus, although they  participate in the critical dynamics at finite wavelengths, 
they decouple at late times. 
Consequently, owing to the dominance of the $O(N)$ charge fluctuations the critical 
dynamics of the multicomponent $O(N)$ theory is 
described  by the dynamical universality class of model G. 

\subsection{Bulk viscosity}

Before summarizing the main result of this work 
we briefly discuss the behavior of the bulk viscosity at the phase transition. 
The properties of the bulk viscosity in a slowly relaxing fluid 
and its possible singular behavior were first addressed 
in Ref.~\cite{Mandelstam} (see also Ref.~\cite{Landau06}). 
The behavior of the bulk viscosity in system with a single component non-conserved 
order parameter was considered in Ref.~\cite{Polyakov}. 
Here, however,  the critical exponent was not evaluated, 
but rather it was guessed based on input from experiment. 

In contrast to the shear viscosity, 
the bulk viscosity can diverge at the critical point in the $O(N)$ model 
depending on the value of $N$. 
For the case of the single component scalar theory in $d=4-\varepsilon$ spatial dimensions, 
the bulk viscosity tends to infinity as 
$\zeta\sim \xi^{z-\alpha/\nu} = \xi^{2}$  (to leading order in $\varepsilon$), 
while for the  multicomponent $N>1$ theory 
the bulk viscosity remains finite 
$\zeta\sim \xi^{0}$. Here $z$ is the dynamical critical exponent, 
which was determined from the slowest mode as a function of $N$.

In order to address the critical behavior of the bulk viscosity, 
the longitudinal component 
of the momentum current has to be considered  in Eq.~(\ref{jEoM}). 
In this case the projection operator on the transverse direction 
is dropped and an additional contribution owing to the bulk viscosity is added on the right hand side of Eq.~(\ref{jEoM}).  
The critical behavior of the bulk viscosity can be deduced  along the lines discussed in Ref.~\cite{Onuki:bulk}, 
where the dynamical critical exponent for the bulk viscosity in model H was computed. 
Also, the QCD critical end point, 
which is theoretically expected to exist at a finite density and temperature in the QCD phase diagram \cite{Asakawa:1989bq}, 
belongs to the universality class of model H \cite{Son:2004iv}. 
Recently the critical dynamics of the QCD critical end point was examined in a comprehensive manner based on DRG 
\cite{Minami:2011un}. 

The results of Ref.~\cite{Onuki:bulk} for the bulk viscosity can be immediately 
generalized to the single component scalar field theory 
since the (non)conservation of the order parameter does not affect 
the result as soon as the dynamical critical exponent is defined.
Consequently, for  $N=1$ the bulk viscosity  diverges at the critical point as $\zeta\sim \xi^{z-\alpha/\nu} $. 
Indeed, the bulk viscosity is given by (see e.g. Refs~\cite{Onuki:bulk,Onuki:book})
\begin{equation}
\zeta = \frac{1}{d^2 T} \lim_{\omega \to 0} \int_0^\infty dt \int d^d x e^{- i \omega t} 
\left\langle \Pi_{ii} (\vec{x},t) \Pi_{jj} (0,0) \right\rangle, 
\label{bulkvisc}
\end{equation}
where $ \Pi_{ij} $ is the stress tensor, 
which can be defined by comparing Eq.~(\ref{jEoM}) 
with the Euler equation $\partial J_i /\partial t = \nabla_j \Pi_{ij}$. 
We are interested in only the dominant singular contribution to Eq.~(\ref{bulkvisc}).  As noted in  Ref.~\cite{Onuki:bulk}, 
it arises from the part of the stress tensor that  is proportional to $\gamma \phi^2 $. 
The integral in  Eq.~(\ref{bulkvisc}) is taken over the domain with characteristic spatial extension of 
order $\xi$ and in the time direction of order $\xi^z$.
Therefore, the dominant singular contribution to the  bulk viscosity reads 
$\zeta \sim \xi^{z-d} \gamma^2 \chi_\phi^2$, 
which reduces to $\zeta\sim \xi^{z-\alpha/\nu} $ 
after substitution of the renormalized quantities for $\gamma$ and $\chi_\phi$. 
In contrast to model H, one should,
however, keep in mind that in this expression $z=2+\alpha/\nu$. 
Thus, extrapolating to $\varepsilon\rightarrow 1$, 
we find that the singularity of the bulk viscosity is given by $\zeta \sim \xi^{2}$, 
while in  model H it is stronger, $\zeta \sim \xi^{2.8}$.
Note, that in both cases  the ratio of the singular part of the  bulk viscosity 
to the relaxation time of the $O(N)$ charge fluctuations, $\tau$, vanishes at the critical point 
as $\zeta/\tau \sim \xi^{-\alpha/\nu}\sim C^{-1}$, in agreement with \cite{Sasaki:2008fg}. 
This is a consequence of the fact that 
the single component scalar field theory 
belongs to the same static universality class as the liquid-gas phase transition. 

For $N>1$ the above discussion does not apply 
because the energy fluctuation decouples from the order parameter in statics, i.e. 
$\gamma^* \to 0$, as we found in Section III. 
Owing to the vanishing mode-mode coupling $g^* \to 0 $ in the long wave limit, 
the critical fluctuations do not couple dynamically to the current $J_i$ either. 

Therefore, the bulk viscosity is finite $\zeta \sim \xi^{0}$ at the critical point. 
In this case, the ratio of the bulk viscosity to 
the relaxation time vanishes
as $\zeta/\tau \sim O(\xi^{-z})$.  

\section{Summary}

In this paper we have evaluated the critical exponents for the dynamics 
of the $O(N)$ scalar field theory with all possible slow modes. 
We showed that for the case of the single component theory 
its dynamical universality class reduces to model C. 
The dynamical critical exponent is given by $z=2+\alpha/\nu$. On the other hand, for the multicomponent theory, the critical dynamics is dominated by 
$O(N)$ charge fluctuations. This drives the critical exponent down to the value $z=d/2$ and the theory belongs to
the dynamic  universality class of model G .
In both cases, $N=1$ and $N>1$, the shear viscosity remains finite at the critical point, 
while the bulk viscosity diverges for $N=1$, and remains finite for $N>1$.

In QCD, the $O(4)$ chiral symmetry in the light quark sector is broken by the finite $u$ and $d$ quark masses. 
For high temperatures and small values of the chemical potential, 
the second-order phase transition is replaced by a crossover. 
Our results imply that 
the singular part of the shear and bulk viscosity remain finite also at the QCD phase transition. 
However, from the present analysis within the DRG, we cannot draw any conclusions on the behavior of the regular   
parts of the viscosities
near a second-order or a crossover transition. 
This problem can only be addressed in more microscopic approaches based on QCD or QCD-like models
\cite{Chen:2007jq,Dobado:2008vt,Chen:2010vg,Hidaka:2009ma,Bluhm:2010qf}, or
within the novel microscopic approach to critical dynamics,  
employing the conjectured gravity dual description of conformal field theories 
\cite{Maeda:2008hn,Buchel:2010gd,Natsuume:2010bs}. 

We acknowledge useful discussions with J. Berges.   
BF acknowledges partial support by the ExtreMe Matter Institute EMMI. 
Work of EN is supported by the Grant-in-Aid for Scientific Research No.~22840031 
and by the EMMI visiting scientist program. 

%

\appendix

\section{Derivation of stochastic equation of motion}
In critical dynamics we are interested only in tracing the evolution of slow modes. 
 From  microscopic point of view even if 
we start with a set of exact equation of motions for the slow modes, 
they would be inevitably affected by all the other degrees of freedom (including fast modes). 
Slow modes also would mix after finite elapse time. 
Therefore, we need a method to extract time evolution of  slow modes only 
from full microscopic equation of motion, 
where the other degrees of freedom are fairly incorporated. 
\subsection{Master equation with projection}
We first derive a master equation, i.e., an equation of motion for the distribution function 
$g_a(t)=\delta\lk A(t)-a\rk=\Pi_l \delta\lk A_l(t)-a_l \rk$, 
which defines a probability distribution 
for the  macroscopic variable $A_l(t)$ to take the value $a_l$ at the time moment $t$. 
We start with Liouville equation 
\beq
\dert A_l(t) = i L A_l(t), 
\eeq
where since we are dealing with slow modes, 
the operator $L$ is supposed to be Poisson bracket with the classical Hamiltonian, 
\beq
i L A_l(t) = \left[ H, A_l(t) \right]_{PB}. 
\eeq

We would like to split $g_a(t)$ into systematic and fluctuating parts.  
	At initial time we start from a state defined by the 
	slow variables. In general there is no a priori rule 
	for the choice of slow variables. The integral of motion are, however, 
	required to be included among slow modes. The slow variables at 
	initial time $A_l(0)$ will be rotated in Hilbert space by 
 the Liouville operator $\exp(i t L)$. This would take 
 $A_l(t)$ out of the subset of slow modes. By the systematic part of  $g_a(t)$ 
 we mean the amount of an overlap  between initial and elapsed distributions at time $t$. 
Therefore. 
it is reasonable to define a projection onto initial state with equilibrium average 
$\langle \cdots\rangle$, 
\beq
Pg_a(t)\equiv \sum_b \langle g_a(t) g_b(0) \rangle g_b(0) 
\eeq
with general properties of the projection operator such as $P P =P$, 
$P+\bar{P}=1$ and $\langle Pg_a(t) \bar{P}g_b(t) \rangle=0$.  
The time evolution of $g_a(t)=Pg_a(t)+\bar{P}g_a(t)$ is given by 
\beq
\dert g_a(x,t)=-\sum_l \frac{\partial}{\partial a_l} \ldk v_l (a) g_a(x,t) \rdk 
+\int db \int ds \langle i L F_a(s) ;b \rangle g_b(x,t-s) +F_a(t), 
\label{dgdt}
\eeq
where 
\beq
 v_l(a) &=& 
\frac{\langle iLA_l(0) g_a(0) \rangle}{\langle g_a(0) \rangle}
\equiv \langle iLA_l(0); a \rangle \nn
F_a(t)
&=& i e^{it \bar{P}L}\bar{P} L g_a(0). 
\eeq
In deriving the above equation, we have used the following decomposition of  Liouville operator:
$e^{itL}=e^{it\bar{P}L}+ i \int_0^t{\rm d}s\, e^{isL} P L e^{i(t-s) \bar{P} L}$, which can be verified 
by taking  time derivative from both sides. 
\subsection{From Master equation to Langevin equation}
One can derive non-linear Langevin equation for $A_i(t)$ 
from Eq.~(\ref{dgdt}) by taking the first moment of the distribution function, 
$A_l(t)= \int {\rm d}a\, a_l g_a(t)$, 
\beq
\dert A_l(t)
&=&
v_l [A(t)]
+\int_0^t ds \langle i L R_l(s); A(t-s)\rangle  +R_l(t), 
\label{eq3_01} \\
&\simeq&
- \frac{\partial \langle \left[  A_l(0), A_m(0) \right]_{PB}; A(t) \rangle}{\partial A_m(t)}  
+\langle \ltk A_l(0), A_m(0) \rtk_{PB}; A(t) \rangle 
\frac{\partial H[A(t)]}{\partial A_m(t)}  \nn
&&-L_{lm}[A(t)]
\frac{\partial H[A(t)] }{\partial A_m(t)} +R_l(t), 
\label{eq3_01mm}
\eeq 
where we introduced the effective Hamiltonian for macroscopic variables $H(a)\equiv -\ln \langle g_a \rangle $  in units of $k_B T=1$, 
and  
\beq
v_l [A(t)] &=& v_l (a)  |_{a=A(t)} \equiv \langle iL A(0) ; a\rangle |_{a=A(t)},  \\
\label{R}
R_l(t)&\equiv& 
e^{it \bar{P}L}\bar{P} \dot{A}_l(0)=i e^{it \bar{P}L}\bar{P} LA_l(0), \\
L_{lm}[ A(t)]&\equiv& \int_0^{\infty} {\rm d}s
\frac{\langle R_l(s)  R_m(0) g_a(0) \rangle }{\langle g_a(0) \rangle}|_{a=A(t)}. 
\eeq
Note that in derivation of Eq.~(\ref{eq3_01mm}) from Eq.~(\ref{eq3_01})
the Markovian approximation for the memory term 
$\int_0^t{\rm d}s \cdots A(t-s) \rightarrow \int_0^\infty{\rm d}s \cdots A(t)$ was applied. 
We also assumed  that the background transport coefficient 
$L_{lm}(A)$ is approximately independent on $A_l(t)$ at late times. 
Owing to properties of Poisson brackets the first term in Eq. (\ref{eq3_01mm}) vanishes in most cases. 

Important point here is that since $R_l(t)\propto \bar{P}$, 
the force $R_l(t)$  is uncorrelated with any macroscopic variables 
by construction 
$\langle G [  A(0) ] R_l(t) \rangle =0$ 
for any arbitrary function $G [ A(0) ]$.  
In this sense the force $R_l(t)$ is a pure random force. 

 The second term in Eq.~(\ref{eq3_01mm}), known also as mode-mode coupling,  describes reversible process, 
and involve non-linear interactions among $A_l(t)$, responsible for critical dynamics. 
\subsection{Application to $O(N)$ model}
Substituting the slow mode candidates in $O(N)$ model and their effective Hamiltonian 
to the above Langevin equation, we obtain 
\beq
\frac{\partial \phi_i}{\partial t} &=& 
-\lambda_0 \lk r_0 \phi_i -\vec{\nabla}^2 \phi_i +u_0\phi_j^2 \phi_i 
+2\gamma_0 \phi_i E-h_i \rk 
- g_0 \lk \vec{\nabla} \phi_i \rk \cdot \lk \vec{J}-\vec{H}\rk \nn
&&+2\tilde{g}_0 \phi_j \lk \chi_Q^{-1}Q_{ij} -\mu_{ij}\rk + \theta_i, \\
\frac{\partial E}{\partial t} &=& 
\Gamma_0\vec{\nabla}^2 \lk C_0^{-1}E-\vec{\nabla}^2E+\gamma_0\phi_i^2 +\beta \rk
- g_0 \lk \vec{\nabla}E \rk \cdot \lk \vec{J}-\vec{H}\rk+ \theta_E, \\
\frac{\partial \vec{J}}{\partial t} 
&=& \mathcal{T}\cdot \ldk 
\eta_0 \vec{\nabla}^2 \lk \vec{J}-\vec{H}\rk
+ g_0 \lk \vec{\nabla} \phi_i \rk \lk
r_0 \phi_i -\vec{\nabla}^2 \phi_i +u_0\phi_j^2 \phi_i+2\gamma_0 \phi_i E -h_i  \rk 
\right. \nn 
&& \left.
+ g_0 \vec{\nabla} E \lk C_0^{-1}E-\vec{\nabla}^2E+\gamma_0\phi_i^2+\beta\rk 
+ g_0 \vec{\nabla} Q_{AB} \lk \chi_Q^{-1} Q_{AB} -\mu_{AB}\rk 
+ \vec{\theta}_J \rdk, \quad \quad \quad \\
\frac{\partial Q_{AB}}{\partial t} &=& 
\Pi_0\vec{\nabla}^2 \lk \chi_Q^{-1}Q_{AB}-\mu_{AB} \rk
- g_0 \lk \vec{\nabla}Q_{AB} \rk \cdot \lk \vec{J}-\vec{H}\rk \nn
&&+\tilde{g}_0 \lk \phi_B h_A-\phi_A h_B 
-\phi_A \vec{\nabla}^2\phi_B
+\phi_B \vec{\nabla}^2\phi_A\rk 
+ \theta_{AB}. 
\eeq 
%
In Fourier space $k \equiv  \{ \omega, \vec{k}\}$, 
the  formal solution is given by 
\beq
\phi_i(k)&=&  G_\phi^0(k) \times \ldk 
\lambda_0 h_i(k) +\theta_i(k)  
-ig_0 \int_1 \phi_i(k_1) \, \vec{k_1} \cdot  \ltk \vec{J}(k-k_1)
-\vec{H}(k-k_1)\rtk \right. \nn
&& 
\left. -\lambda_0u_0 \int_{12} \phi(k_1)_j\phi(k_2)_j\phi_i(k-k_1-k_2) 
-2\lambda_0\gamma_0 \int_{1} \phi_i(k_1)E(k-k_1)  \rdk \nn
&&+2\tilde{g}_0 \int_1 \phi_j(k_1) 
\ldk \chi_Q^{-1} Q_{ij}(k-k_1)-\mu_{ij}(k-k_1)\rdk, \\
E(k)&=& 
G_E^0(k) \times \ldk -\Gamma_0 \vk^2 \beta(k) +\xi(k)
-\Gamma_0 \gamma_0 \vk^2 \int_1 \phi_i(k_1)\phi_i(k-k_1) \right. \nn 
&&\left. -ig_0 \int_1  E(k_1) \vk_1 \cdot \ltk \vec{J}(k-k_1)-\vec{H}(k-k_1)\rtk  
\rdk, \\
\vec{J}(k)
&=&G_J^0(k) \cdot {\mathcal T}_k \cdot \ldk 
\eta_0  \vec{k}^2 \vec{H}(k)+\vec{\zeta}(k)\right. \nn 
&& 
+ig_0\int_1 
\vec{k_1} \phi_i(k_1) \ltk \chi_\phi^{-1}(\vk-\vk_1)  
\phi_i(k-k_1)-h_i(k-k_1)+
2\gamma_0\int_{12}\phi_i(k_2) E(k-k_1-k_2) \rtk  \nn 
&& 
+i g_0 u_0 \int_{123} \vec{k_1} \phi_i(k_1)\phi_j(k_2)\phi_j(k_3)
\phi_i(k-k_1-k_2-k_3) \nn
&&
+\left.  ig_0\int_1 
\vec{k_1} E(k_1) \ltk 
\chi_E^{-1}(\vk-\vk_1) E(k-k_1) + \beta(k-k_1) 
+\gamma_0 \int_2 \phi_i(k_2)\phi_i(k-k_1-k_2) \rtk \rdk \nn
&&+ig_0 \int_1 Q_{AB}(k_1) \vk_1
\ldk \chi_Q^{-1} Q_{AB}(k-k_1)-\mu_{AB}(k-k_1)\rdk, \\
Q_{AB}(k)&=&
G_Q^0(k) \ldk \Pi_0 \vk^2 \mu_{AB}(k) +\theta_Q(k)
-ig_0 \int_1 Q_{AB}(k_1) \vk_1 \cdot \ltk \vec{J}(k-k_1)-\vec{H}(k-k_1) \rtk \right. 
\nn
&&
+\tilde{g}_0 \int_1  \ltk \phi_B(k_1) h_A(k-k_1)-\phi_A(k_1) h_B(k-k_1) 
\right. \nn
&&\left. \left. 
+\phi_A(k_1) \phi_B(k-k_1) \lk \vk^2-2\vk\cdot \vk_1\rk\rtk \rdk, 
\eeq
where $\chi_\phi^{-1}(\vk)=r_0 +\vk^2$ the inverse of static susceptibility,
$\int_{12\cdots m}\equiv \int \Pi_{n=1}^m
{\rm d}\omega_n{\rm d}^dk_n/(2\pi)^{d+1}$ and 
$\lk {\mathcal T}_k\rk_{ij} = \delta_{ij}-k_i k_j/\vec{k}^2$. 
Bare propagators read 
\beq
G_J^0(k) \cdot {\mathcal T}_k 
&=&\frac{1}{-i\omega +\eta \vec{k}^2} {\mathcal T}_k, \\
G_E^0(k)
&=&\frac{1}{-i\omega +\Gamma_0 \vec{k}^2 \lk C_0^{-1} +\vk^2 \rk},\\
G_\phi^0(k)
&=&\frac{1}{-i\omega +\lambda_0 \lk r_0+\vec{k}^2\rk},\\
G_Q^0(k)
&=&\frac{1}{-i\omega +\Pi_0 \chi_Q^{-1}\vk^2}. 
\eeq
Note that $G_\phi^0(k)$ and $G_Q^0(k)$ are of diagonal form in $O(N)$ space. 

The noise-noise correlation functions satisfy the  fluctuation-dispersion relations, 
\beq
\langle \theta_i(k) \theta_j(k')\rangle 
&=&2 \lambda_0 \, \delta_{ij} \delta\lk \vec{k}+\vec{k'}\rk,\\ 
\langle \xi(k) \xi(k')\rangle 
&=&2 \Gamma_0 \vec{k}^2 \, \delta\lk \vec{k}+\vec{k'}\rk,\\ 
\langle \zeta_i(k) \zeta_j(k')\rangle 
&=&2 \eta_0 \vec{k}^2 \, \delta_{ij} \delta\lk \vec{k}+\vec{k'}\rk. 
\eeq

The renormalized transport coefficients are determined from the response functions, 
\beq
\phi_i(\omega, \vk)
&=&\chi_{\phi_i}(\omega, \vk)_{ij} h_j(\omega, \vk), \\
E(\omega, \vk)
&=&-\chi_{E}(\omega, \vk)\beta(\omega, \vk), \\
J_i(\omega, \vk)
&=&{\bf \chi}_{J}(\omega, \vk) \lk{\mathcal T}_{\vk}\rk_{ij} H_j(\omega, \vk). 
\eeq


\section{Response function}
The response functions of   slow modes are obtained 
from  a set of stochastic equation of motions.  
The old-fashioned perturbation method, i.e., 
iteration of formal solution for different orders of  interaction terms 
and taking average over noises in energy-momentum space, 
systematically generates loop corrections to response functions.  
In this article we perform only leading order calculations.  In order to 
proceed with the calculations beyond the leading order, 
it is preferable to use the alternative field-theoretical 
approach (already mentioned in the text). 

\subsection{Order parameter relaxation constant}
The response function for order parameter fluctuations is given as follows: 
\beq
\chi_\phi(k)
&=&
\left \langle \left. \frac{\delta \phi_i(k)}{\delta h_i(k)}\right|_{h_i \rightarrow 0}
\right\rangle_{\theta}
=
G_\phi(k) \ldk  \lambda +\Sigma_\phi(k) \rdk, 
\eeq
where loop corrections are accounted for in $\Sigma_\phi(k)$. 
%
\begin{figure}[H]
  \begin{center}
      \resizebox{120mm}{!}{\includegraphics{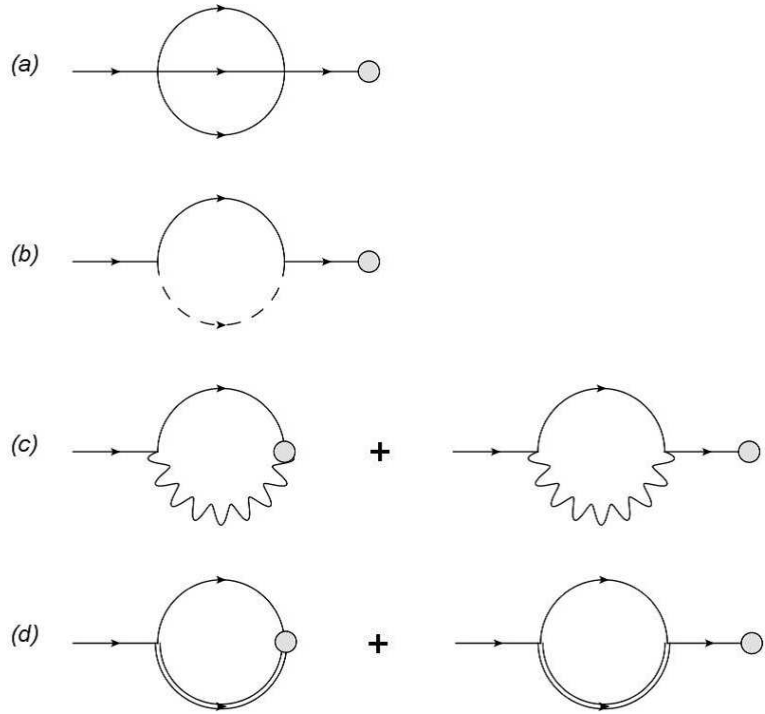}} 
    \caption{Leading order contributions to $\chi_\phi$. 
Solid line represents propagator of order parameter fluctuations, 
dashed  --  energy, wavy -- transverse momentum, and double -- $O(N)$ charge fluctuations, respectively.
The solid circle indicates external field.
}
    \label{selfphi1}
  \end{center}
\end{figure}
%
The diagrams for  the leading order contributions are shown in Fig.~\ref{selfphi1}, 
and the corresponding equations are given by  
\beq  
\Sigma^a_\phi(k)
&=& 2 \lk N+2 \rk \lk \lambda u\rk^2 \int_{\vec{1},\vec{2}}
\frac{3r+\vk_1^2+\vk_2^2+\lk \vk-\vk_1-\vk_2\rk^2}
{\ldk r+\lk \vk-\vk_1-\vk_2\rk^2\rdk \ldk r+\vk_1^2\rdk \ldk r+\vk_2^2\rdk}\nn
&&\times \frac{1}{-i\omega 
+\lambda \ldk 3r+\lk \vk-\vk_1-\vk_2\rk^2+\vk_1^2+\vk_2^2\rdk}
\\
\Sigma^b_\phi(k)
&=&
\ldk \lk -2\lambda \gamma \rk^2 
\int_{1} 2 \Gamma \lk \vk-\vk_1 \rk^2 G_E(k_1-k) G_E(k-k_1) G_\phi(k_1) \right. \nn
&&+ \left. \lk -2\lambda \gamma \rk \lk -\Gamma \gamma \rk 
\int_{1} \lk \vk-\vk_1 \rk^2 2 \lambda G_\phi(k_1) G_\phi(-k_1) G_E(k-k_1) \rdk 
G_\phi(k) \lambda \nn 
&=& 4 G_\phi(k) \lambda^2 \gamma^2 C  \int_{\vec{1}} 
\frac{1}
{r+\vk_+^2} 
\frac{\lambda \lk r+\vk_+^2\rk+\Gamma/C \vk_-^2}
{-i\omega +\lambda \lk r+\vk_+^2\rk+\Gamma/C \vk_-^2}, \\
\Sigma^c_\phi(k)
&=&
2\lambda_0 g_0^2 \int_{1} G_\phi^0(k_1) G_\phi^0(-k_1) 
G_J(k-k_1) \vec{k}_1\cdot {\mathcal T}_{\vec{k}-\vec{k}_1}\cdot \vec{k}_1 
+ \cdots \nn
&=& \ldk 1-\lambda \lk r+\vk^2\rk G_\phi(k) \rdk g^2 \int_{\vec{1}} 
\frac{1}{r+\vk_+^2}
\frac{\vec{k}\cdot {\mathcal T}_{\vec{k}_-}\cdot \vec{k}}
{-i\omega +\lambda\lk r+\vk_+^2\rk +\eta\vk_-^2}, \\
\Sigma^d_\phi(k)
&=&
2\chi_Q^{-1} \tilde{g}^2 \lambda \sum_j \int_{1}G_{Q_{ij}}(k-k_1) 
G_{\phi_{j}}(k_1) G_{\phi_{j}}(-k_1) + \cdots \nn
&=&
\frac{\tilde{g}^2 (N-1)}{\chi_Q} 
\ldk 1-\lambda \lk r+\vk^2\rk G_\phi(k)\rdk \int_{\vec{1}}
\frac{1}{r+\vk_+^2}
\frac{1}{-i\omega+\lambda(r+\vk_+^2)+\Pi\chi_Q^{-1}\vk_-^2}. 
\eeq
The renormalized order parameter relaxation constant to leading order is given by 
\beq
\lambda_{\rm ren}^{-1}=
\left. \frac{\partial \chi_\phi^{-1}}{\partial \lk -i\omega\rk}\right|_{k\rightarrow 0}=
\lambda^{-1} \ldk 1 
+\frac{4\gamma^2 C \lambda}{\lambda+\Gamma/C}\Omega_4 \ln b
-\frac{\tilde{g}^2 \lk N-1\rk}{\lambda \chi_Q\lk \lambda+\Pi/\chi_Q\rk}\Omega_4 \ln b \rdk . 
\eeq
where $\vk_\pm=\vk_1\pm \vk/2$. 
We evaluate the above equation near $d=4$ and at the critical point, 
where a renormalized mass goes like $r \sim 0$ 
(note that non-trivial fixed point $r^*$ can be adjusted to $0$). 
Since $\Sigma_\phi^b \sim \vk^2$, 
$\lambda$ acquires no mode-mode coupling contribution of order of $g^2$ 
at zero  momentum. 

In the following, 
we take the same procedure to obtain 
the other transport coefficients renormalized to the leading order. 
\subsection{Energy diffusion constant}
\begin{figure}[H]
  \begin{center}
      \resizebox{120mm}{!}{\includegraphics{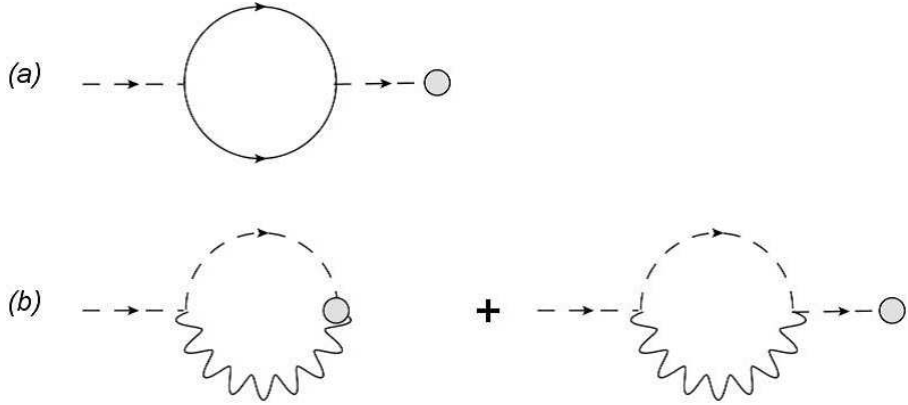}} 
    \caption{Leading order loop corrections to $\chi_E$.}
    \label{fig2}
  \end{center}
\end{figure}
%
The response function for energy fluctuation is given by 
\beq
\chi_E(k)
&=&
-\left \langle \left. \frac{\delta E(k)}{\delta \beta(k)}\right|_{\beta \rightarrow 0}
\right\rangle
=G_E(k) \ldk  \Gamma \vk^2 +\Sigma_E(k) \rdk. 
\eeq
Leading order contributions $\Sigma_E$ are depicted in Fig.~2, and are given by 
\beq
\Sigma^a_E(k) 
&=& G_E(k)\sum_i \lk 2 \lambda \gamma \Gamma\vk^2\rk^2 \int_{1} 
G_\phi(k_1)\ldk G_\phi(-k_1)G_\phi(k-k_1)+G_\phi(k-k_1)G_\phi(k_1-k)\rdk\nn
&=& 
G_E(k)N  2 \lambda \lk \gamma \Gamma\vk^2\rk^2\int_{\vec{1}} 
\frac{2 \lk r+\vk_1^2+\vk^2/4\rk }{\lk r+\vk_-^2\rk \lk r+\vk_+^2\rk}
\frac{1}{-i\omega +\lambda\lk r+\vk_-^2\rk+\lambda\lk r+\vk_+^2\rk}\\
\Sigma^b_E(k) 
&=&  -g^2 \int_{1} 2\Gamma \vk_1^2 G_E(k_1) G_E(-k_1) 
G_J(k-k_1) \vec{k}_1\cdot {\mathcal T}_{\vec{k}-\vec{k}_1}\cdot \lk -\vec{k}_1 \rk \nn
&+& g^2 \int_{1} \vk_1\cdot {\mathcal T}_{\vk -\vk_1}\cdot \vk 
(-2\eta)\lk \vk-\vk_1\rk^2 
G_E(k_1) G_J(k_1-k)G_J(k-k_1)\nn 
&=& g^2 \ldk C - \Gamma \vk^2 G_E(k)\rdk \int_{\vec{1}} 
\frac{\vec{k}\cdot {\mathcal T}_{\vec{k}_-}\cdot \vec{k}}
{-i\omega +\Gamma C^{-1}\vk_+^2 +\eta \vk_-^2}, 
\eeq
where $\vk_\pm=\vk_1\pm \vk/2$. 

Evaluating equations above near $d=4$ and at the critical point: 
\beq
\Gamma_{\rm ren}^{-1}
&=&\left. \vk^2\frac{\partial \chi_E^{-1}}{\partial \lk -i\omega\rk}\right|_{k\rightarrow 0}
\simeq \left. \frac{1}{\Gamma} 
\ldk 1-G_E^{-1}\frac{{\Sigma_E^b}'}{\Gamma \vk^2}\rdk \right| \nn
&=& \frac{1}{\Gamma} 
\ldk 1-\frac{3}{4}\frac{g^2}{\Gamma C^{-1}\lk \Gamma C^{-1}+\eta \rk}\Omega_4\frac{\Lambda^2}{2}
\lk 1-b^{-2}\rk\rdk .
\eeq

\subsection{Shear viscosity}
\begin{figure}[H]
  \begin{center}
      \resizebox{120mm}{!}{\includegraphics{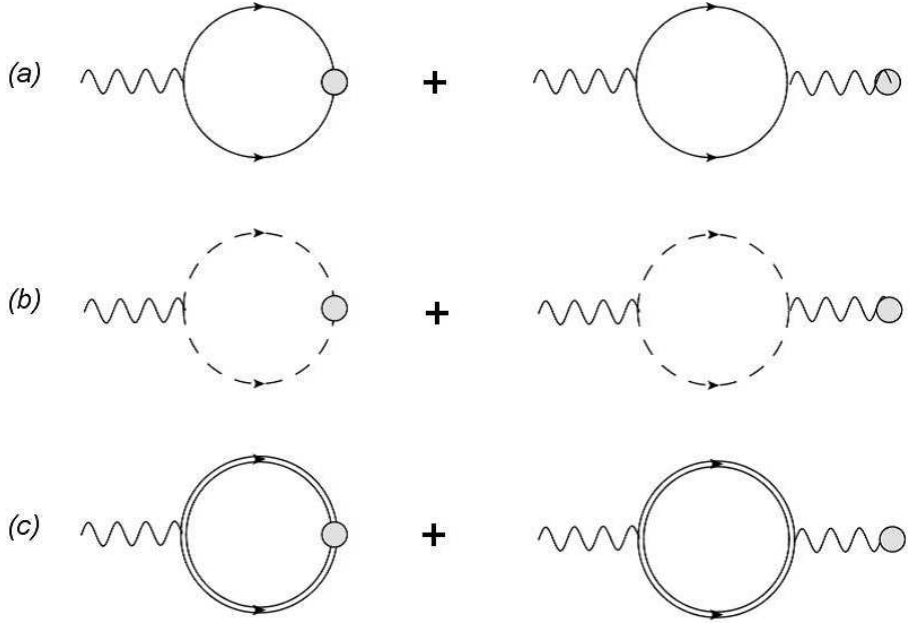}} 
    \caption{Loop corrections to $\chi_J$.}
    \label{fig3}
  \end{center}
\end{figure}
%
The response function for the  transverse momentum is of a tensor form 
because of the projection operator ${\mathcal T}$, 
with leading order corrections $\Sigma_J$ shown in Fig.~3, 
\beq
\chi_J(k) \lk {\mathcal T}_\vk\rk_{ij}
&=& 
\left \langle \left. \frac{\delta J_i(k)}{\delta H_j(k)}\right|_{H\rightarrow 0}
\right\rangle
=
G_J(k) \lk {\mathcal T}_\vk\rk_{ij'}\cdot 
\ldk  \eta \vec{k}^2 +\Sigma_J(k) \rdk_{j'j}.  
\eeq
\beq
\Sigma_J^a(k) 
&=& 
-2 \lambda g^2 \int_{1} {\mathcal T}_{\vec{k}}\cdot\vec{k}_1 G_\phi(k_1)  
\ldk r+\lk \vk-\vk_1 \rk^2  \rdk G_\phi(k-k_1) 
\vk_1\cdot {\mathcal T}_\vk  \nn 
&&\times
\ldk  
  G_\phi(k_1-k) 
 -G_\phi(-k_1) \rdk + \cdots\nn 
&=&
2g^2 \ldk 1-\eta \vk^2 G_J(k) \rdk \int_{\vec{1}}
\frac{ \vk\cdot \vk_1}{r+\vk_-^2}
\frac{{\mathcal T}_\vk\cdot\vk_1 \vec{k}_1\cdot{\mathcal T}_{\vec{k}}}
{-i\omega +\lambda_0 \lk r+\vk_+^2\rk
+\lambda_0 \lk r+\vk_-^2\rk}, \\
\Sigma_J^b(k)
&=&
\Sigma_J^c(k) = 0, 
\eeq
where $\vk_1 \rightarrow -\lk \vk_1-\vk/2\rk$ to reach the last equality 
in $\Sigma_J^a$. 
The transverse part $\Sigma_J$ is extracted by projection 
(${\rm Tr} {\mathcal T}_\vk=d-1$), 
\beq
\Sigma_J\lk k \rk=
\frac{2g_0^2}{d-1} \ldk 1-\eta \vk^2 G_J(k) \rdk \int_{\vec{1}}
\frac{ \vk\cdot \vk_1}{r+\vk_-^2}
\frac{\vk_1\cdot{\mathcal T}_\vk \cdot\vec{k}_1}
{-i\omega +\lambda_0 \lk r+\vk_+^2\rk
+\lambda_0 \lk r+\vk_-^2\rk}. 
\eeq
For $r\sim 0$ 
\beq
\left. G_J^{-1}\frac{\Sigma_J'\lk k \rk}{\eta\vk^2} \right|_{k\rightarrow 0}
&\simeq&\left. 
G_J\frac{2g_0^2}{d-1} \int_{\vec{1}}
\frac{ \vk\cdot \vk_1}{\vk_-^2}
\frac{\vk_1\cdot{\mathcal T}_\vk \cdot\vec{k}_1}
{\lambda_0 \lk \vk_+^2+\vk_-^2\rk}\right|\nn 
&\simeq&
\frac{2g_0^2}{3\eta} \Omega_4 \int_{\Lambda/b}^\Lambda {\rm d}k_1 k_1^3
\frac{\left\langle \lk \hat{k}\cdot \hat{k}_1\rk^2 
\ldk 1-\lk\hat{k}\cdot \hat{k}_1\rk^2\rdk \right\rangle_4}
{2\lambda_0 k_1^2}\nn 
&=&
\Omega_4\frac{g^2}{24\lambda\eta} \frac{\Lambda^2}{2}\lk 1-\frac{1}{b^2}\rk, 
\eeq
where $\langle \cdots \rangle_4$ implies taking an average on the solid angle at $d=4$: 
a $d$-dimensional angle average of an even power of one momentum component is defined by  
\beq
\langle k_i^{2n} \rangle_d &\equiv &
\frac{\int {\rm d}^dk \, k_i^{2n} \delta\lk \vk^2-1\rk }
{\int {\rm d}^dk \delta\lk \vk^2-1\rk }
=
\frac{1\cdot 3\cdots (2n-1)}{2^n}\frac{\lk d/2-1\rk!}{\lk n+d/2-1\rk!}. 
\eeq
For the present use 
$\langle k_i^{2} \rangle_4=\frac{1}{4}$ and $\langle k_i^{4} \rangle_4=\frac{1}{8}$. 
The renormalized $\eta$ eventually reads 
\beq
\eta_{\rm ren}^{-1}
=\left. \vk^2\frac{\partial \chi_J^{-1}}{\partial \lk -i\omega\rk}\right|_{k\rightarrow 0}
=\eta^{-1}\ldk 1- \frac{g^2}{24\lambda\eta} \Omega_4 \frac{\Lambda^2}{2}\lk 1
-b^{-2}\rk\rdk. 
\eeq

\subsection{$O(N)$ charges diffusion constant}
The response function for the $O(N)$ charges diffusion constant is defined by
\beq
\chi_Q(k) 
&=& 
\left \langle \left. \frac{\delta Q_{ij}(k)}{\delta \mu_{ij}(k)}
\right|_{\mu\rightarrow 0}\right\rangle
=G_Q(k) \ldk  \Pi \vec{k}^2 +\Sigma_Q(k) \rdk , 
\eeq
\begin{figure}[H]
 \begin{center}
      \resizebox{120mm}{!}{\includegraphics{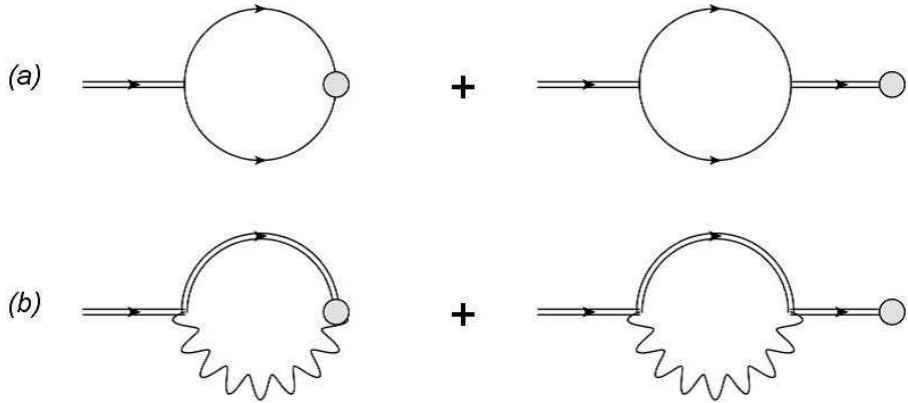}} 
    \caption{Loop contributions to $\chi_Q$.}
    \label{fig4}
  \end{center}
\end{figure}
%
%
where different contributions depicted in Fig.~\ref{fig4} read  
\beq
\Sigma_Q^a(k) 
&=& 
-2 \lambda \tilde{g}^2 \int_{1} \lk \vk^2 -2\vk\cdot\vk_1\rk 
G_\phi(k_1) \ldk G_\phi(-k_1)  G_\phi(k-k_1) - G_\phi(k_1-k)  G_\phi(k-k_1)\rdk
+ \cdots \nn 
&=&
\tilde{g}^2 \ldk 1- \Pi \chi^{-1} \vk^2 G_Q(k) \rdk \int_{\vec{1}} 
\frac{4\lk \vk \cdot \vk_1\rk^2}{\lk r+\vk_-^2 \rk \lk r+\vk_+^2 \rk }
\frac{1}
{-i\omega +\lambda \lk r+\vk_-^2 \rk+\lambda \lk r+\vk_+^2 \rk}, \\
\Sigma_Q^b(k) 
&=& 
2 \Pi g^2 \int_{1} \vk_1^2 \vk_1\cdot {\mathcal T}_{k-k}\cdot\vk_1 
G_Q(k_1) G_Q(-k_1)  G_J(k-k_1) 
\ldk \delta_{ik}\delta_{jl}- \delta_{il}\delta_{jk} (-1)\rdk + \cdots \nn 
&=&
2g^2 \int_{\vec{1}}
\frac{\vec{k}\cdot{\mathcal T}_{k_-}\cdot\vk}
{-i\omega +\Pi\vk_+^2
+\eta \vk_-^2}. 
\eeq
The renormalized $O(N)$ charge diffusion constant is given by 
\beq
\Pi_{\rm ren}^{-1}&=&\left. \vk^2 \frac{\partial \chi_Q^{-1}}{\partial(-i\omega)}
\right|_{k \rightarrow 0} 
=\left. \Pi_{0}^{-1} 
\ldk {G_Q^{-1}}' \lk 1-\frac{\Sigma_Q}{\Pi_0 \vk^2}\rk 
+ G_Q^{-1} \lk 1-\frac{\Sigma_Q}{\Pi_0 \vk^2}\rk' \rdk\right| \nn
&\simeq& \Pi^{-1} \ldk 1
-\frac{3g^2}{4\lk\Pi+\eta\rk\Pi}\Omega_4\Lambda^2\lk 1-b^{-2}\rk
-\frac{\tilde{g}^2}{2\lambda\Pi}\Omega_4\ln b
\rdk, 
\eeq
where we have taken the solid angle average: 
\beq
\left. \frac{\Sigma_Q}{\Pi_0 \vk^2}\right|
&=&\frac{2g^2}{\lk\Pi+\eta\rk\Pi_0}\int_{\vec{1}}
\frac{\hat{k}\cdot{\mathcal T}_{k_1}\cdot\hat{k}}{\vk_1^2}
\simeq
\frac{3g^2}{2\lk\Pi+\eta\rk\Pi}
\Omega_4\frac{\Lambda^2\lk 1-b^{-2}\rk}{2}, \\ 
\left.G_Q^{-1}\frac{\Sigma_Q'}{\Pi_0 \vk^2}\right|
&=& \left. \chi_Q^{-1} \tilde{g}^2 G_Q 
\int_{\vec{1}} 
\frac{4\lk \vk \cdot \vk_1\rk^2}{\lk r+\vk_-^2 \rk \lk r+\vk_+^2 \rk }
\frac{1}
{-i\omega +\lambda \lk r+\vk_-^2 \rk+\lambda \lk r+\vk_+^2 \rk}\right| \nn
&=& \frac{2\tilde{g}^2}{\lambda_0\Pi_0}
\int_{\vec{1}} 
\frac{\lk \hat{k} \cdot \hat{k}_1\rk^2}{\vk_1^4 }
\simeq\frac{\tilde{g}^2}{2\lambda\Pi}\Omega_4\ln b. 
\eeq


\appendix


\begin{thebibliography}{99}

\bibitem{Ackermann:2000tr}
  K.~H.~Ackermann {\it et al.}  [STAR Collaboration],
  Phys.\ Rev.\ Lett.\  {\bf 86}, 402 (2001)
  [arXiv:nucl-ex/0009011].

\bibitem{Adler:2003kt}
  S.~S.~Adler {\it et al.}  [PHENIX Collaboration],
  Phys.\ Rev.\ Lett.\  {\bf 91}, 182301 (2003)
  [arXiv:nucl-ex/0305013].

\bibitem{Adams:2003am}
  J.~Adams {\it et al.}  [STAR Collaboration],
  Phys.\ Rev.\ Lett.\  {\bf 92}, 052302 (2004)
  [arXiv:nucl-ex/0306007].


\bibitem{Teaney:2003kp}
  D.~Teaney,
  Phys.\ Rev.\  C {\bf 68}, 034913 (2003)
  [arXiv:nucl-th/0301099].

\bibitem{Shuryak:2003xe}
  E.~Shuryak,
  Prog.\ Part.\ Nucl.\ Phys.\  {\bf 53}, 273 (2004)
  [arXiv:hep-ph/0312227].

\bibitem{Romatschke:2007mq}
  P.~Romatschke and U.~Romatschke,
  Phys.\ Rev.\ Lett.\  {\bf 99}, 172301 (2007)
  [arXiv:0706.1522 [nucl-th]].

\bibitem{Gavin:1985ph}
  S.~Gavin,
  Nucl.\ Phys.\  {\bf A435}, 826-843 (1985).

\bibitem{Prakash:1993bt}
  M.~Prakash, M.~Prakash, R.~Venugopalan and G.~Welke,
  Phys.\ Rept.\  {\bf 227}, 321 (1993).


\bibitem{Davesne:1995ms}
  D.~Davesne,
  Phys.\ Rev.\  C {\bf 53}, 3069 (1996).

\bibitem{Dobado:2003wr}
  A.~Dobado and F.~J.~Llanes-Estrada,
  Phys.\ Rev.\  D {\bf 69}, 116004 (2004)
  [arXiv:hep-ph/0309324].

\bibitem{Chen:2006iga}
  J.~W.~Chen and E.~Nakano,
  Phys.\ Lett.\  B {\bf 647}, 371 (2007)
  [arXiv:hep-ph/0604138].

\bibitem{Chen:2007xe}
  J.~W.~Chen, Y.~H.~Li, Y.~F.~Liu and E.~Nakano,
  Phys.\ Rev.\  D {\bf 76}, 114011 (2007)
  [arXiv:hep-ph/0703230].


\bibitem{Chen:2007kx}
  J.~W.~Chen and J.~Wang,
  Phys.\ Rev.\  C {\bf 79}, 044913 (2009)
  [arXiv:0711.4824 [hep-ph]].

\bibitem{Sasaki:2008fg}
  C.~Sasaki and K.~Redlich,
  Phys.\ Rev.\  C {\bf 79}, 055207 (2009)
  [arXiv:0806.4745 [hep-ph]].


\bibitem{Arnold:2000dr}
  P.~B.~Arnold, G.~D.~Moore and L.~G.~Yaffe,
  JHEP {\bf 0011}, 001 (2000)
  [arXiv:hep-ph/0010177].

\bibitem{Arnold:2003zc}
  P.~B.~Arnold, G.~D.~Moore and L.~G.~Yaffe,
  JHEP {\bf 0305}, 051 (2003)
  [arXiv:hep-ph/0302165].

\bibitem{Xu:2007jv}
  Z.~Xu, C.~Greiner and H.~Stocker,
  Phys.\ Rev.\ Lett.\  {\bf 101}, 082302 (2008)
  [arXiv:0711.0961 [nucl-th]].


\bibitem{Chen:2009sm}
  J.~W.~Chen, H.~Dong, K.~Ohnishi and Q.~Wang,
  Phys.\ Lett.\  B {\bf 685}, 277 (2010)
  [arXiv:0907.2486 [nucl-th]].

\bibitem{Niemi:2011ix}
  H.~Niemi, G.~S.~Denicol, P.~Huovinen, E.~Molnar and D.~H.~Rischke,
  arXiv:1101.2442 [nucl-th].




\bibitem{Karsch:1986cq}
  F.~Karsch and H.~W.~Wyld,
  Phys.\ Rev.\  D {\bf 35}, 2518 (1987).

\bibitem{Nakamura:2004sy}
  A.~Nakamura and S.~Sakai,
  Phys.\ Rev.\ Lett.\  {\bf 94}, 072305 (2005)
  [arXiv:hep-lat/0406009].


\bibitem{Meyer:2007ic}
  H.~B.~Meyer,
  Phys.\ Rev.\  D {\bf 76}, 101701 (2007)
  [arXiv:0704.1801 [hep-lat]].

\bibitem{Meyer:2007dy}
  H.~B.~Meyer,
  Phys.\ Rev.\ Lett.\  {\bf 100}, 162001 (2008)
  [arXiv:0710.3717 [hep-lat]].

\bibitem{Karsch:2007jc}
  F.~Karsch, D.~Kharzeev and K.~Tuchin,
  Phys.\ Lett.\  B {\bf 663}, 217 (2008)
  [arXiv:0711.0914 [hep-ph]].

\bibitem{Kharzeev:2007wb}
  D.~Kharzeev and K.~Tuchin,
  JHEP {\bf 0809}, 093 (2008)
  [arXiv:0705.4280 [hep-ph]].

\bibitem{Huebner:2008as}
  K.~Huebner, F.~Karsch and C.~Pica,
  Phys.\ Rev.\  D {\bf 78} (2008) 094501
  [arXiv:0808.1127 [hep-lat]].


\bibitem{Pisarski:1983ms}
  R.~D.~Pisarski and F.~Wilczek,
  Phys.\ Rev.\  D {\bf 29}, 338 (1984).




\bibitem{Hosoya:1983id}
  A.~Hosoya, M.~A.~Sakagami and M.~Takao,
  Annals Phys.\  {\bf 154}, 229 (1984).

\bibitem{Jeon:1994if}
  S.~Jeon,
  Phys.\ Rev.\  D {\bf 52}, 3591 (1995)
  [arXiv:hep-ph/9409250].

\bibitem{Jeon:1995zm}
  S.~Jeon and L.~G.~Yaffe,
  Phys.\ Rev.\  D {\bf 53}, 5799 (1996)
  [arXiv:hep-ph/9512263].


\bibitem{Aarts:2003bk}
  G.~Aarts and J.~M.~Martinez Resco,
  Phys.\ Rev.\  D {\bf 68}, 085009 (2003)
  [arXiv:hep-ph/0303216].

\bibitem{Aarts:2004sd}
  G.~Aarts and J.~M.~Martinez Resco,
  JHEP {\bf 0402}, 061 (2004)
  [arXiv:hep-ph/0402192].

\bibitem{York:2008rr} 
  M.~A.~York and G.~D.~Moore,
  Phys.\ Rev.\ D {\bf 79}, 054011 (2009)
  [arXiv:0811.0729 [hep-ph]].

\bibitem{Carrington:2009kh} 
  M.~E.~Carrington and E.~Kovalchuk,
  Phys.\ Rev.\ D {\bf 81}, 065017 (2010)
  [arXiv:0912.3149 [hep-ph]].

\bibitem{Hidaka:2010gh}
  Y.~Hidaka and T.~Kunihiro,
  arXiv:1009.5154 [hep-ph].

\bibitem{Carrington:2004tm} 
  M.~E.~Carrington and S.~Mrowczynski,
  Phys.\ Rev.\ D {\bf 71}, 065007 (2005)
  [hep-ph/0406097].

\bibitem{Berges:2005md} 
  J.~Berges and S.~Borsanyi,
  Phys.\ Rev.\ D {\bf 74}, 045022 (2006)
  [hep-ph/0512155].




\bibitem{Chen:2007jq}
  J.~W.~Chen, M.~Huang, Y.~H.~Li, E.~Nakano and D.~L.~Yang,
  Phys.\ Lett.\  B {\bf 670}, 18 (2008)
  [arXiv:0709.3434 [hep-ph]].


\bibitem{Chen:2010vg} 
  J.~-W.~Chen, M.~Huang, C.~-T.~Hsieh and H.~-H.~Lin,
  Phys.\ Rev.\ D {\bf 83}, 115006 (2011)
  [arXiv:1010.3121 [hep-ph]].

\bibitem{Schafer:2009dj}
  T.~Schafer, D.~Teaney,
  Rept.\ Prog.\ Phys.\  {\bf 72}, 126001 (2009).
  [arXiv:0904.3107 [hep-ph]]. 





\bibitem{Berges:2000ew}
  J.~Berges, N.~Tetradis and C.~Wetterich,
  Phys.\ Rept.\  {\bf 363}, 223 (2002)
  [arXiv:hep-ph/0005122].


\bibitem{Janssen}
 H.~K.~Janssen, Z. Phys. B {\bf 23}, 377 (1976).

\bibitem{DeDominicis:1977fw}
  C.~De Dominicis and L.~Peliti,
  Phys.\ Rev.\  B {\bf 18}, 353 (1978).

\bibitem{Canet:2006xu}
  L.~Canet and H.~Chate,
  J.\ Phys.\ A {\bf 40}, 1937 (2007)
  [arXiv:cond-mat/0610468].



\bibitem{Hohenberg:1977ym}
  P.~C.~Hohenberg and B.~I.~Halperin,
  Rev.\ Mod.\ Phys.\  {\bf 49}, 435 (1977).


\bibitem{Folk:2006ve}
  R.~Folk and H.~G.~Moser,
  J.\ Phys.\ A  {\bf 39}, R207 (2006).


\bibitem{Fisher:1974uq}
  M.~E.~Fisher,
  Rev.\ Mod.\ Phys.\  {\bf 46}, 597 (1974)
  [Erratum-ibid.\  {\bf 47}, 543 (1975)].

\bibitem{Wilson:1973jj}
  K.~G.~Wilson and J.~B.~Kogut,
  Phys.\ Rept.\  {\bf 12}, 75 (1974).




\bibitem{zwa1}
R. Zwanzig, Phys. Rev. 124 (1961), 983. 

\bibitem{Mori1}
H.~Mori and H.~Fujisaka, Prog. Theor. Phys. Vol. 49 No. 3 (1973) pp. 764-775

\bibitem{Mori2}
H.~Mori, Prog. Theor. Phys. Vol. 49 No. 5 (1973) pp. 1516-1545


\bibitem{Fix}
M.~Fixman, J. Chem. Phys. 36, (1962), 310. 

\bibitem{Kawa}
 K.~Kawasaki, Ann. Phys. 61  (1970), pp. 1-56. 

\bibitem{zwa}
  R.~Zwanzig,
  ``Nonequilibrium statistical mechanics,''
  Oxford University press, (2001). 

\bibitem{Berges:2009jz}
  J.~Berges, S.~Schlichting and D.~Sexty,
  arXiv:0912.3135 [hep-lat].


\bibitem{Halperin:1974zz}
  B.~I.~Halperin, P.~C.~Hohenberg and S.~k.~Ma,
  Phys.\ Rev.\  B {\bf 10}, 139 (1974).


\bibitem{Mandelstam}
L.I. Mandelshtam and M.A. Leontovich, Zh. Eksp. Teor. Fiz. {\bf  7 } 438 (1937).

\bibitem{Landau06}
L.D.~Landau  and E.M. Lifshitz, ``Fluid mechanics'', Pergamon Press  (1979), p. 304.

\bibitem{Polyakov}
A.M. Polyakov, 
Sov. Phys. JETP {\bf 30}  1164 (1969).

\bibitem{Onuki:bulk}
  A.~Onuki,
  Phys.\ Rev.\  E {\bf 55}, 403 (1997).

\bibitem{Asakawa:1989bq}
  M.~Asakawa, K.~Yazaki,
  Nucl.\ Phys.\  {\bf A504}, 668-684 (1989).

\bibitem{Son:2004iv}
  D.~T.~Son, M.~A.~Stephanov,
  Phys.\ Rev.\  {\bf D70}, 056001 (2004).
  [hep-ph/0401052].


\bibitem{Minami:2011un}
  Y.~Minami,
  Phys.\ Rev.\  {\bf D83}, 094019 (2011).
  [arXiv:1102.5485 [hep-ph]].

\bibitem{Onuki:book}
  A.~Onuki, ``Phase transition dynamics'', Cambridge University press, (2002).




\bibitem{Dobado:2008vt}
  A.~Dobado, F.~J.~Llanes-Estrada and J.~M.~Torres-Rincon,
  Phys.\ Rev.\  D {\bf 79}, 014002 (2009)
  [arXiv:0803.3275 [hep-ph]].

\bibitem{Hidaka:2009ma}
  Y.~Hidaka and R.~D.~Pisarski,
  Phys.\ Rev.\  D {\bf 81}, 076002 (2010)
  [arXiv:0912.0940 [hep-ph]].

\bibitem{Bluhm:2010qf}
  M.~Bluhm, B.~Kampfer and K.~Redlich,
  arXiv:1011.5634 [hep-ph].


\bibitem{Maeda:2008hn}
  K.~Maeda, M.~Natsuume and T.~Okamura,
  Phys.\ Rev.\  D {\bf 78}, 106007 (2008)
  [arXiv:0809.4074 [hep-th]].

\bibitem{Buchel:2010gd}
  A.~Buchel,
  Nucl.\ Phys.\  B {\bf 841}, 59 (2010)
  [arXiv:1005.0819 [hep-th]].


\bibitem{Natsuume:2010bs}
  M.~Natsuume and T.~Okamura,
  arXiv:1012.0575 [hep-th].




\end{thebibliography}
\end{document}